\documentclass[a4paper,fleqn,usenatbib]{mnras}
\usepackage{newtxtext,newtxmath}

\usepackage[T1]{fontenc}
\usepackage{ae,aecompl}

\usepackage{graphicx}	
\usepackage{amsmath}	
\usepackage{longtable,lscape}
\usepackage{multicol}
\usepackage{multirow}
\usepackage{natbib}
\usepackage{txfonts}
\usepackage{hyperref}
\usepackage{caption}
\usepackage{tabularx}
\usepackage[utf8]{inputenc}
\usepackage{array}
\usepackage{float}
\usepackage{epstopdf}
\usepackage{xcolor}

\def\m2s2{\,m$^{2}$\,s$^{-2}$} 

\title[COPAINS Survey]{Results from The COPAINS Pilot Survey:
four new brown dwarfs and a high companion detection rate for accelerating stars 
}

\author[M. Bonavita et al.]{M. Bonavita$^{1,2,3}$\thanks{E-mail: mariangela.bonavita@open.ac.uk}, 
C. Fontanive$^{4}$, 
R. Gratton$^{3}$,
K. Mu{\v z}i{\'c}$^{5}$,
S. Desidera$^{3}$,
D. Mesa$^{3}$,
B. Biller$^{2}$,
\newauthor 
A. Scholz$^{6}$,
A. Sozzetti$^{7}$,
V. Squicciarini$^{3,8}$
\\
$^{1}$School of Physical Sciences, The Open University, Walton Hall, Milton Keynes, MK7 6AA\\
$^{2}$SUPA, Institute for Astronomy, University of Edinburgh, Blackford Hill, Edinburgh EH9 3HJ, UK\\
$^{3}$INAF Osservatorio Astronomico di Padova, Vicolo dell'Osservatorio 5,  35121 Padova, ITALY\\
$^{4}$Center for Space and Habitability, University of Bern, Bern 3012, Switzerland\\
$^{5}$CENTRA, Faculdade de Ci\^{e}ncias, Universidade de Lisboa, Ed. C8, Campo Grande, P-1749-016 Lisboa, Portugal\\
$^{6}$SUPA, School of Physics \& Astronomy, University of St Andrews, North Haugh, St Andrews, KY16 9SS, UK\\
$^{7}$INAF - Osservatorio Astrofisico di Torino, Via Osservatorio 20, 10025, Pino Torinese, Italy\\
$^{8}$Department of Physics and Astronomy {\it Galileo Galilei}, University of Padova, Via dell’Osservatorio 3, I-35122 Padova, Italy
}

\date{Accepted 2022 April 25. Received 2022 April 21; in original form 2022 April 08}

\begin{document}
\label{firstpage}
\pagerange{\pageref{firstpage}--\pageref{lastpage}}
\maketitle

\begin{abstract}
The last decade of direct imaging (DI) searches for sub-stellar companions has uncovered a widely diverse sample that challenges the current formation models, while highlighting the intrinsically low occurrence rate of wide companions, especially at the lower end of the mass distribution. 
These results clearly show how blind surveys, crucial to constrain the underlying planet and sub-stellar companion population, are not an efficient way to increase the sample of DI companions. It is therefore becoming clear that efficient target selection methods are essential to ensure a larger number of detections. 
We present the results of the COPAINS Survey conducted with SPHERE/VLT, searching for sub-stellar companions to stars showing significant proper motion differences ($\Delta\mu$) between different astrometric catalogues. We observed twenty-five stars and detected ten companions, including four new brown dwarfs: HIP 21152 B, HIP 29724 B, HD 60584 B and HIP 63734 B.  
Our results clearly demonstrates how astrometric signatures, in the past only giving access to stellar companions, can now thanks to Gaia reveal companions well in the sub-stellar regime.
We also introduce FORECAST (Finely Optimised REtrieval of Companions of Accelerating STars), a tool which allows to check the agreement between position and mass of the detected companions with the measured $\Delta\mu$. Given the agreement between the values of the masses of the new sub-stellar companions from the photometry with the model-independent ones obtained with FORECAST, the results of COPAINS represent a significant increase of the number of potential benchmarks for brown dwarf and planet formation and evolution theories.

\end{abstract}




\section{Introduction}
\label{sec:intro}
Direct imaging (DI) is the only detection method that provides observations of an exoplanet or brown dwarf (BD) itself, as it captures the thermal emission of self-luminous companions. With the unique opportunity to obtain photometric and spectroscopic observations of substellar objects, this detection method allows for a direct probe of cold companions atmospheres. DI is also necessary to study the outer regions of planetary systems, that cannot be probed by other detection methods.\\
Despite the remarkable efforts that have been invested in the development of new observing technologies and image processing techniques, and a steady increase in the census of wide-orbit companions, only a handful of systems below the deuterium-burning limit have been uncovered around stars in DI programs, and the occurrences of wide companions appear to be intrinsically low \citep{Biller07,Biller13,Lafreniere07,NielsenClose10,Vigan12,Vigan2017,Vigan21,Rameau13,Galicher16,Nielsen19}.
In order to empirically constrain the formation, evolution, and atmospheric properties of both isolated and bound sub-stellar companions, we need to uncover a substantial population of these objects, and measure their fundamental properties, such as the effective temperature and mass. However, even when a comprehensive view and an extensive spectro-photometric characterisation is possible, imaging surveys still only provide measurements of an object's luminosity. Mass estimates for imaged planets and brown dwarfs therefore rely entirely on evolutionary models, which currently carry high uncertainties, particularly at young to intermediate ages. An independent determination of masses from dynamical arguments is therefore crucial to overcome the large uncertainties introduced by evolutionary models, and in turn refine the theories. 
Furthermore, a sample of benchmark objects should ideally span a wide range of properties (e.g., spectral types, masses, ages). As direct imaging is more amenable to very young systems, where low-mass companions are still bright, the majority of the scarce sample of such benchmark objects in the planetary regime orbit relatively young stars (few tens to hundred Myr). On the contrary, most brown dwarf companions with well-defined dynamical masses orbit older hosts with field ages of several Gyrs (see Fig.\,1). As known bound sub-stellar companions are even rarer in associations such as Hyades 
\citep[$\sim$650 Myr][]{martin2018}, the intermediate age regime remains relatively unexplored, and theoretical models are hence particularly poorly constrained at these ages. 
Our understanding of the origins and atmospheres of these objects thus remains severely limited by the small number of known systems. In particular, the sparse sample of directly-imaged companions show a large diversity of spectro-photometric characteristics and orbital configurations, and remain challenging to grasp as populations \citep{Bowler16}. Larger numbers of detections are hence essential to enable a better characterisation and understanding of the wide-orbit companion population, and obtain a clearer picture of their formation patterns.\\
In \citet{Fontanive2019}, we presented a new tool \texttt{COPAINS} (Code for Orbital Parametrisation of Astrometrically Inferred New Systems), developed to identify previously undiscovered companions detectable via DI, based on changes in stellar proper motions across multiple astrometric catalogues. A significant proper motion difference ($\Delta\mu$) between two catalogues for a given star is a good indication of the presence of a perturbing body. For systems showing significant differences between proper motions measured over a long time baseline (e.g., Tycho-2, or Tycho {\it Gaia}Astrometric Solution - TGAS; \citealt{tycho-2,TGAS}), and catalogues that provide short-term proper motions (e.g., Hipparcos, {\it Gaia}DR2; \citealt{Hipparcos, GaiaDR2}), the tool allows for the computation of secondary mass and separation pairs compatible with the observed trend, marginalised over all possible orbital phases and eccentricities. The resulting solutions are based entirely on dynamical arguments, although a dependence on the adopted (usually model-derived) stellar mass remains in the obtained secondary masses. Compared to the expected sensitivity of an imaging instrument, these predictions can then be used to select the most promising targets for DI searches of low-mass companions.\\
The use of such informed selection processes had already proven to be effective in the stellar regime \citep{MakarovKaplan05,Tokovinin13,Bowler21,Steiger21}, and recently led to the discovery of new brown dwarf companions based on the astrometric signatures induced on their host stars \citep{Currie20,Chilcote21}. Such astrometric systems are particularly valuable, as the combination of relative astrometry from DI information with absolute astrometry from the primary's astrometric signature offers a remarkable opportunity to refine orbital constraints and measure dynamical masses \citep{CalissendorffJanson18,Snellen18,Brandt19,Dupuy19,Grandjean19,Maire20,Nielsen20,Drimmel2021}. Precise orbital elements for the population of wide-orbit companions can provide key insights into formation mechanisms (e.g., \citealp{Bowler20}).
Furthermore, model-independent mass measurements for brown dwarfs and giant planets are especially important to bypass the use of mass estimates from theoretical models, which typically carry large uncertainties, both due to the difficulties of determining system ages, and to the systematic uncertainties of the evolutionary and atmosphere models which are particularly pronounced at the lowest masses and youngest ages. Increasing the pool of systems amenable to dynamical mass measurements will therefore be essential to help calibrate theoretical models for substellar objects.\\
In this paper, we present the results of a pilot survey conducted with the SPHERE instrument \citep{sphere}, an extreme adaptive optics facility at the ESO's Very Large Telescope (VLT), which employed the \texttt{COPAINS} tool for informed target selection. We describe the sample and selection method in Section~\ref{sec:sample}. The observations and data reduction are presented in Section~\ref{sec:observations}. The survey results are reported in Section~\ref{sec:results} and discussed in Section~\ref{sec:discussion}.

\section{Sample Properties}
\label{sec:sample}

\subsection{Target Selection}
\label{sec:selection}

\subsubsection{Initial Target List}
To select targets for DI campaigns, we searched different catalogues containing proper motions, using as an input a list of known, relatively young, sources, from which the targets protected by the SPHERE Guaranteed Time Observations (GTO) were removed. For the sources with both long- and short-term proper motion information, we first selected stars showing a difference larger than $3 \sigma$ between two catalogues, in either of the proper motion components. 
The data presented here were obtained during three ESO periods (P100, P102, and P104)\footnote{ESO programme IDs 0100.C-0646, 0102.C-0506, 0104.C-0965 }, with several differences in target selection procedure, as a result of the different $Gaia$ data releases available at the time of each selection. The relevant information about the two initial selections are listed here.\\
 
 \noindent {\bf Selection 1 (P100 and P102)}
 \begin{itemize}
    \item Input list: An extensive compilation of about 900 nearby young stars \citep{Desidera15}.
    \item Long-term proper motion catalogue: Tycho-2 \citep{tycho-2}, with a proper motion baseline of $\sim$100\,yr.
    \item Short-term proper motion catalogue: The Tycho-{\it Gaia}Astrometric Solution (TGAS; \citealp{TGAS}), part of $Gaia$ DR1 \citep{GaiaDR1}, with a proper motion baseline of $\sim$25\,yr.
 \end{itemize}
 
 \noindent {\bf Selection 2 (P104)}
 \begin{itemize}
    \item Input list:  The same compilation as above \citep{Desidera15}, complemented by bona-fide members of nearby young moving groups \citep{gagne2018,gagnefaherty2018}. In total, there are about 2200 unique objects in this input list.
    \item Long-term proper motion catalogue: TGAS, with a proper motion baseline of $\sim$25\,yr.
    \item Short-term proper motion catalogue: $Gaia$ DR2 \citep{GaiaDR2}, with a proper motion baseline of $\sim$1.5\,yr.
 \end{itemize} 
 \noindent Following the selection of these so-called $\Delta \mu$ candidates, stars with known companions from visual and/or radial velocity observations were removed from the list. A star was removed if satisfying any of the following criteria:
 \begin{itemize}
     \item Multiplicity flag (MultiFlag) equal to C or O in the Hipparcos catalogue \citep{Hipparcos};
     \item Star appears in the Catalog of Components of Double \& Multiple stars (CCDM; \citealt{ccdm});
     \item Star appears in The ninth catalogue of spectroscopic binary orbits (S$_B^9$; \citealt{SBcatalogue});
     \item Star has a sub-stellar companion listed in The Extrasolar Planets Encyclopaedia\footnote{\url{http://exoplanet.eu/}};
     \item A suffix in the name of the star indicating that it is part of a binary system, or an unresolved binary (e.g. A, B, AB).
     \item For the Selection 2, we also excluded stars with another source closer than $5'$ on the sky, and sharing the same parallax (within 3$\sigma$), in order to exclude obvious stellar binaries from the survey and correctly focus on sub-stellar companions.  For this, we only used the sources with a relatively small parallax uncertainties ($\varpi$/ $\sigma_\varpi < 0.2$).
 \end{itemize}
 
 \begin{figure*}
    \centering
    \includegraphics[width=0.45\textwidth]{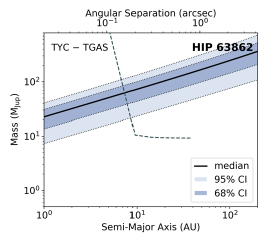}   
    \includegraphics[width=0.45\textwidth]{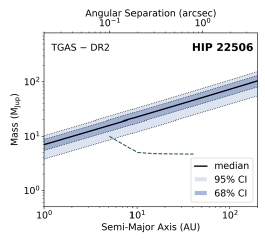}
    \caption{Examples of the solutions computed with \texttt{COPAINS} for the observed astrometric trends of one target selected through the Selection~1 process (using Tycho-2 and TGAS; left) and one target from the Selection~2 process (using TGAS and {\it Gaia}DR2; right). The solid black lines corresponds to the median curves of solutions, and the dark and light shaded areas represent the 1 and 2-$\sigma$ confidence intervals, respectively. 
    The dashed grey lines show the expected detection limit used for the survey selection, computed as detailed in the text.
    }
    \label{fig:COPAINS_trend_main}
\end{figure*}

\subsubsection{COPAINS Selection}\label{sec:copains}
After removing known binaries as detailed above, we further retained only the targets for which the \texttt{COPAINS} tool \citep{Fontanive2019} returned that the objects causing the observed astrometric trends could be sub-stellar, if in the parameter space detectable with SPHERE (i.e., within the instrument field-of-view and above the expected sensitivity of observations). \\
\texttt{COPAINS} provides a good indication of the region of the parameter space in which a hidden companion responsible for an observed astrometric offset may be. Based on the formalism from \citet{MakarovKaplan05}, the approach assumes that long-term proper motion measurements are representative of a system's centre-of-mass motion, and that short-term measurements correspond to the instantaneous reflex motion of the host star. For a measured $\Delta\mu$ value, the code allows us to evaluate the possible companion mass and separation pairs compatible with the astrometric data, for a given distance, stellar mass (see Section~\ref{sec:mstar}), and eccentricity distribution, while assuming face-on orbital inclinations (see \citealp{Fontanive2019} for details). We adopted a Gaussian eccentricity distribution centred around $e$=0 with a width of 0.3 \citep{Bonavita2013}, and the astrometric catalogues listed above were used as long- and short-term proper motion values. \\
Figure~\ref{fig:COPAINS_trend_main} shows examples of the resulting trends computed with \texttt{COPAINS} for one target from P100 (left) and one target from P104 (right). 
In each case, the estimated solutions were compared to the expected sensitivity limit of SPHERE-IFS, shown as a red solid line in Figure~\ref{fig:COPAINS_trend_main}. 
The IFS limits were obtained following the approach described by \citet{mesa2021} and converted to minimum mass limits using the models from \cite{baraffe2015} and the values of the age and stellar mass available at the time of the selection. 
Only promising targets, where the intersection of the detection limits and computed regions suggested possible sub-stellar companions detectable in our survey, were kept for the final sample, based on a visual analysis of the obtained plots. 
Finally, only the targets not previously observed with SPHERE, and observable in the relevant ESO period were kept. 
The final target list consisting of 25 stars observed with SPHERE is given is Table~\ref{tab:master}, while Table \ref{tab:deltamu} lists the values of the proper motions from Tycho-II, TGAS, Gaia DR2 and EDR3 and the resulting $\Delta \mu$.

\subsection{Stellar Ages}
\label{sec:ages}
A full revision of the stellar ages was performed for all the targets following the approaches described in \citet{Desidera15} and \cite{Desidera2021}, considering a variety of indicators and also performing a check for membership to groups using the BANYAN $\Sigma$ on-line tool \footnote{\url{http://www.exoplanetes.umontreal.ca/banyan/banyansigma.php}} \citep{gagne2018}.
Most of the targets were found to be field objects, while still compatible with a relatively young age, or stars with ambiguous membership. In these cases, we considered indirect age indicators such as the equivalent width of 6708\AA\ Lithium doublet, rotation period, X-ray emission, chromospheric activity, taking as reference the empirical sequences of members of groups and clusters \citep[e.g., ][]{Desidera15} and isochrone fitting. Several of our targets have ages between Hyades and Pleiades. In this range, we took advantage of the recently derived rotation sequence for Group X at an age of 300 Myr \citep{messina2022}. When applicable, we considered the age indicators for physical companions outside the SPHERE field of view and, for close binaries, we deblended photometric colors for binarity, to improve the reliability of the derived ages.
For the few target found to be belonging to young moving groups, we adopted the values of the ages presented in \citet{bonavita2016} and \citet{Desidera2021}, mostly based on \citet{bell2015}.\\
All targets were found to be younger than 1~Gyr, except for GJ~3346 for which the age is likely to be closer to 5~Gyr as discussed in detail in \cite{bonavita2020b}. The age determination process for each target is discussed in Appendix~\ref{app:targets}. 

\subsection{Stellar Masses} 
\label{sec:mstar}
The masses for all the stars in the sample were also revised, using the Manifold Age Determination for Young Stars \citep[\textsc{madys}, Squicciarini \& Bonavita in preparation; see][for a description of the tool]{sq2021} and the updated values of the stellar ages. 
\textsc{Madys} retrieved and cross-matched photometry from {\it Gaia} EDR3 \citep{GaiaEDR3} and 2MASS \citep{2mass} for all our targets and then applied a correction for interstellar extinction by integrating along the line of sight the 3D extinction map by \citep{leike20}; the derived A(G) were turned into the photometric band of interest using a total-to-selective absorption ratio $R$=3.16 and extinction coefficients $A_\lambda$ from \cite{wang19}.
The derived absolute magnitudes were then compared with a grid of isochrones with an age range based on the minimum and maximum age values included in Table~\ref{tab:master} to yield a mass estimate. \textsc{Madys} can use several available grids, but in this instance the PARSEC isochrones \citep{marigo17} were used, due to their large dynamical range spanning the entire stellar regime. A constant solar metallicity, appropriate for most nearby star-forming regions, was assumed \citep{dorazi11} for all targets except for HIP 21152 and HIP 21317, for which we assumed [Fe/H]=+0.13 based on their membership to the Hyades (see Appendix~\ref{app:targets} for details). For each star, a sample of mass estimates was constructed by computing the best-fit mass at different ages within the given age range; its median was taken as the final mass estimate, while the reported errors represent the 16th and the 84 percentile; photometric uncertainties were naturally propagated on the final result via a Monte Carlo approach, i.e. by randomly varying, in a Gaussian fashion, photometric data according to their uncertainties while building the sample of mass estimates.

\begin{figure*}
    \includegraphics[width=0.45\textwidth]{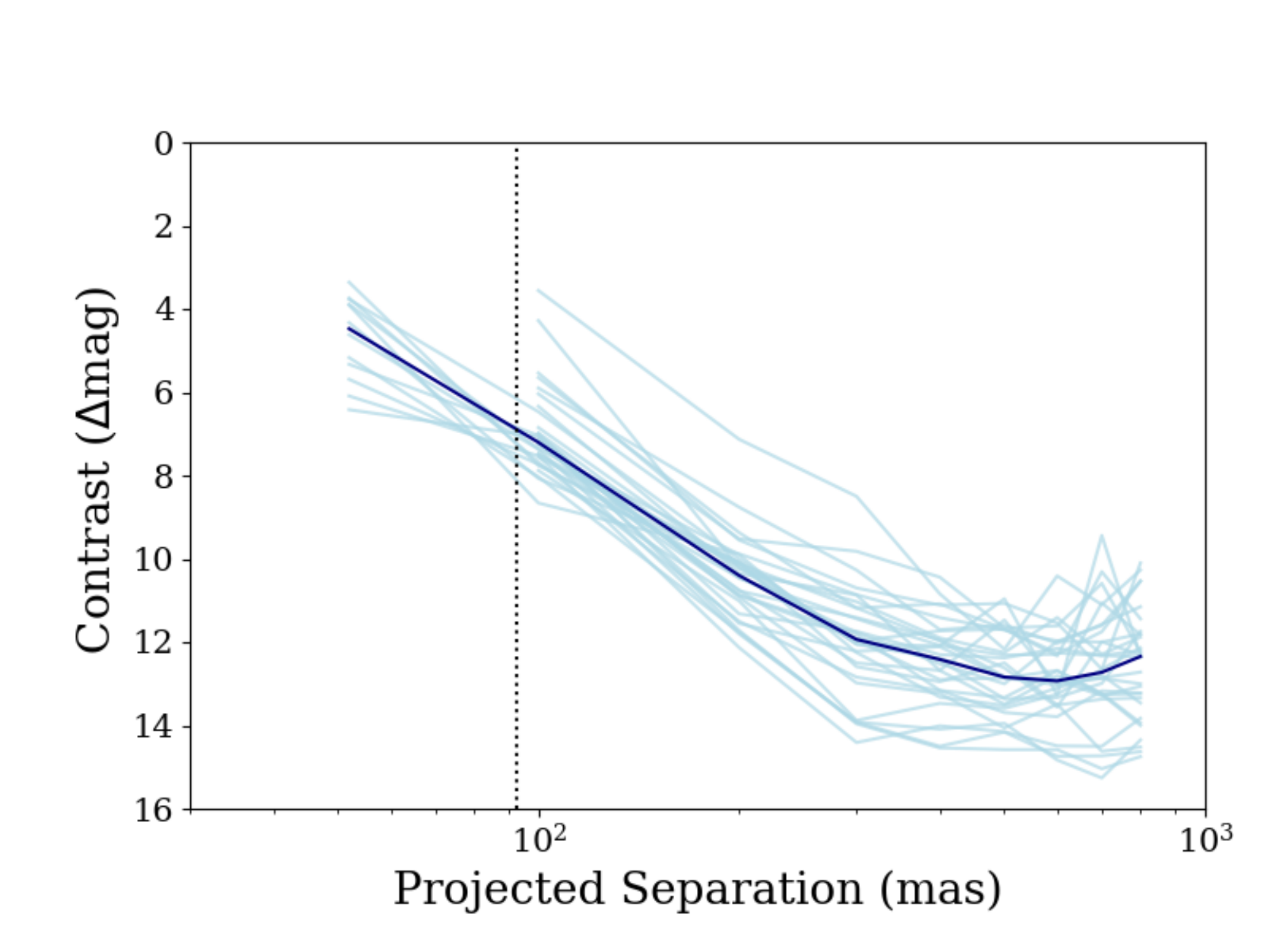}
    \includegraphics[width=0.45\textwidth]{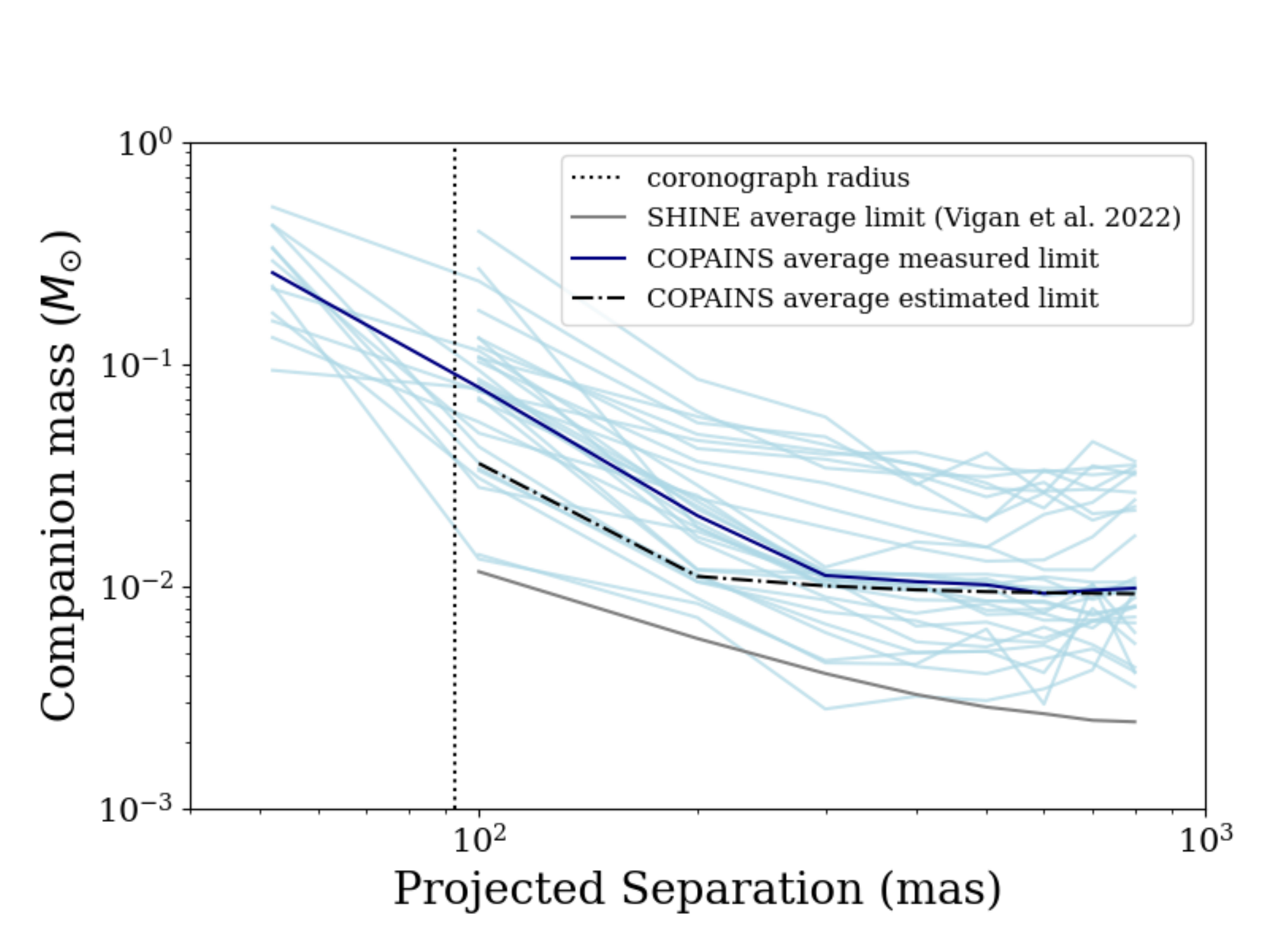}
    \caption{IFS detection limits expressed in contrast (left) and minimum companion mass (right) vs projected separation, for all the targets in our sample. The dashed vertical line marks the coronographic radius. Note that objects with multiple epochs will appear more than once. The COND models \citep{baraffe2003} were used for the magnitude to mass conversion, using the adopted age from Table~\ref{tab:master}.}
    \label{fig:mass_lim}
\end{figure*}

\section{Observations and data reduction} 
\label{sec:observations}
All observations were performed with VLT/SPHERE \citep{sphere} with the two Near Infra-Red (NIR) subsystems, IFS \citep{ifs} and IRDIS \citep{irdis} observing in parallel (IRDIFS Mode), with IRDIS in dual-band imaging mode \citep[DBI;][]{Vigan2010}.
For all targets we used the IRDIFS-EXT mode, which enables covering the $Y$-, $J$-, $H$-, and $K$-band in a single observation, which is meant to provide a high-level of spectral content for subsequent analyses.\\
A summary of the observing parameters and conditions is given in Table~\ref{tab:obslog}. 
The observing sequence adopted was similar to those designed for the SHINE Guaranteed time survey \citep[see e.g.][]{chauvin2017} and consisted of: 
\begin{itemize}
    \item One PSF sub-sequence composed of a series of off-axis unsaturated images obtained with an offset of $\sim$0.4$''$ relative to the coronagraph center (produced by the Tip-Tilt mirror). A neutral density filter was used to avoid saturation\footnote{\url{www.eso.org/sci/facilities/paranal/instruments/sphere/inst/filters.html}} and the AO visible tip-tilt and high-order loops were closed to obtain a diffraction-limited PSF.
    \item A {\it star center} coronagraphic observation with four symmetric satellite spots, created by introducing a periodic modulation on the deformable mirror \citep[see][for details]{langlois2013}, in order to enable an accurate determination of the star position behind the coronagraphic mask for the following deep coronagraphic sequence. 
    \item The deep coronagraphic sub-sequence, for which we used here the smallest apodized Lyot coronagraph (ALC-YH-S) with a focal-plane mask of 185~mas in diameter.
    \item A new star center sequence, a new PSF registration, as well as a short sky observing sequence for fine correction of the hot pixel variation during the night. 
\end{itemize}
IRDIS and IFS data sets were reduced using the SPHERE Data Reduction and Handling (DRH) automated pipeline \citep{Pavlov2008} at the SPHERE Data Center \citep[SPHERE-DC, see][]{Delorme2017} to correct for each data cube for bad pixels, dark current, flat field and sky background. After combining all data cubes with an adequate calculation of the parallactic angle for each individual frame of the deep coronagraphic sequence, all frames are shifted at the position of the stellar centroid calculated from the initial star center position.
In order to calibrate the IRDIS and IFS data sets on sky, we used images of the astrometric reference field 47 Tuc observed with SPHERE at a date close to our observations. The plate scale and true north values used are based on the long-term analysis of the GTO astrometric calibration described by \cite{maire2016}.

\subsection{Detection Limits}
\label{sec:limits}
In order to evaluate our sensitivity to stellar companions, we determined detection limits for point sources. We used the standard procedure to derive detection limits outside the coronagraphic field masks that makes use of the SPECAL software as described in \citet{Galicher2018} and used in the F150 survey \citep{Langlois2021}. The detection limits considered here were obtained using the Template Locally Optimised Combination of Images \citep[TLOCI][]{tloci} for IRDIS and the ASDI-PCA \citep[Angular Spectral Differential Imaging with PCA][]{Galicher2018} for IFS. \\
Contrast limits for the individual data sets are shown in the left panel of Fig.~\ref{fig:mass_lim} and reported in Table~\ref{tab:ifs_ccurves}.
The corresponding values of the minimum companion mass limits, obtained using the evolutionary models from \cite{baraffe2015} for the magnitude to mass conversion, are shown in the right panel Fig.~\ref{fig:mass_lim}. \\
The achieved average limits appear to be significantly worse than the average of the expected limits (dashed-dotted line in Fig.~\ref{fig:mass_lim}) used for the target selection. This is most likely due to the fact that, since our program was executed in service mode and as filler, most of our targets were observed in sub-optimal conditions and with very small field rotation, with a strong negative effect on the quality of the high contrast imaging performances, especially at short separations. This is also confirmed by the fact that the average limit achieved for the first 150 targets of the SHINE survey (gray solid line, from \cite{Vigan21}), where all target were observed in the best possible conditions, is instead much better than both the measured and estimated COPAINS limits. 

\begin{figure*}
    \centering
    \includegraphics[width=\textwidth]{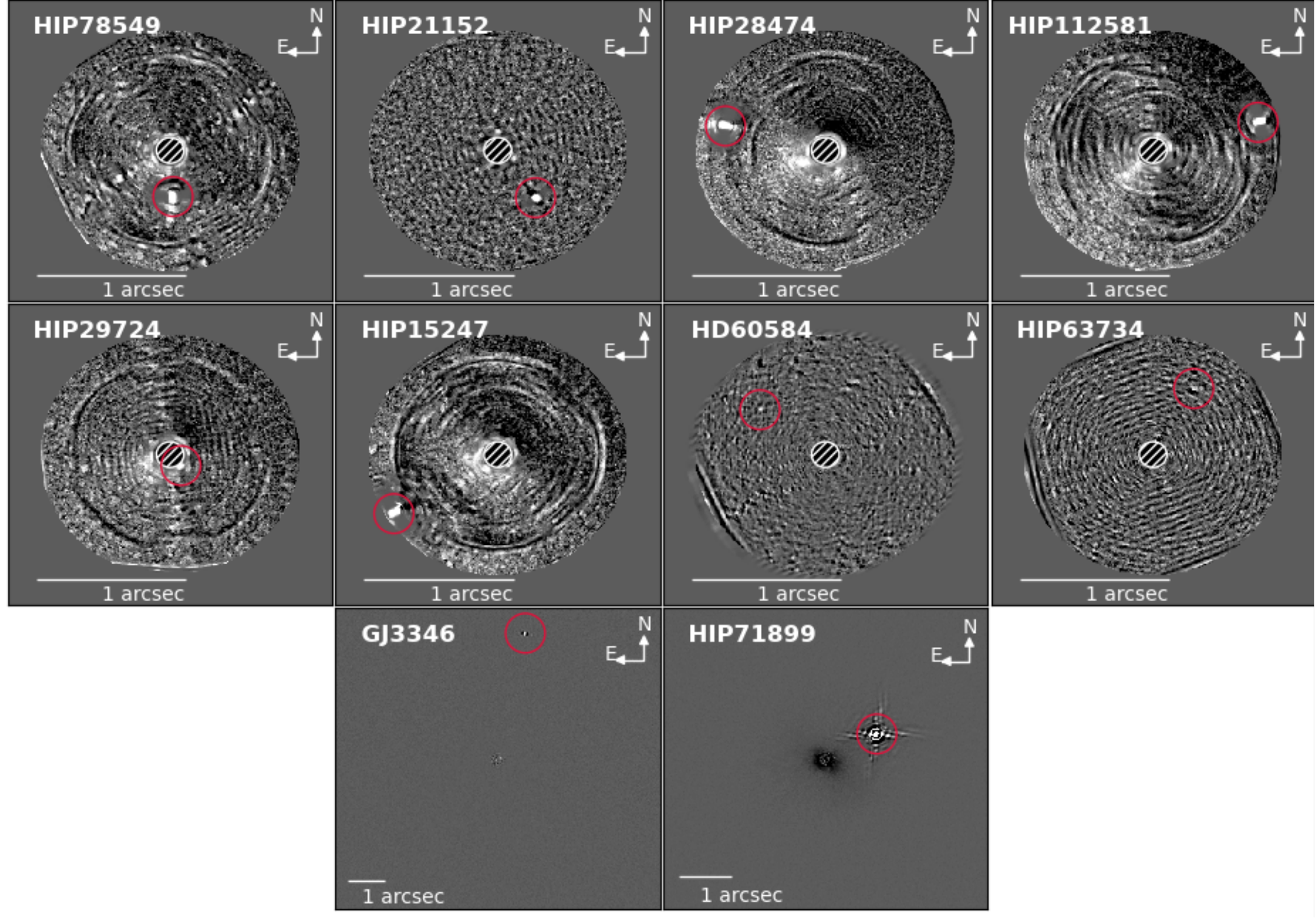}
    \caption{Detected candidate companions in the IFS (top 2 rows) and IRDIS (bottom row) field of view. The red circle marks the position of the candidate. }
    \label{fig:gallery1}
\end{figure*}

\section{Results}
\label{sec:results}
We detected a total of 14 candidate companions, 4 of which were found to be background sources thanks to additional epochs available in the literature or obtained in our program (HIP 63862, HD 57852, HIP 33690). The 10 co-moving companions are shown in Fig.~\ref{fig:gallery1}.
Eight of the co-moving companions have separations below 0.9 arcseconds, and are therefore in the IFS field of view, while the remaining 2 were only observed with IRDIS. 
Five are new discoveries, including a white dwarf companion at $\sim3.6^{\prime\prime}$ from GJ~3346 (already presented in \cite{bonavita2020b}) and four new sub-stellar companions: HIP~21152~B\footnote{HIP 21152~B was independently discovered as part of two other surveys targeting accelerating stars, as detailed in \cite{Kuzuhara:2022arXiv} and Franson et al. 2022 (in preparation). Both works include an in-depth characterisation of the system, the latter also including a full spectral and orbital analysis combining all available data sets.}
,  HIP~29724~B, HD~60584~B and HIP~63734~B. Their properties are discussed in details in Sec.~\ref{sec:hd28736}, \ref{sec:hip29724} and \ref{sec:tyc6539-HIP6734}.

\subsection{SPHERE astrometry and photometry}
\label{sec:sphere_astroph}
The astrometry and photometry measurements from all our SPHERE observations, are listed in Table~\ref{tab:photoastro}. For each epoch we report the projected separation and position angle, and the contrast (expressed as apparent magnitude difference) in the IFS $Y$ and $J$ filters, as well as the IRDIS $K_1$ and $K_2$ for the IRDIFS-EXT observations\footnote{see \url{www.eso.org/sci/facilities/paranal/instruments/sphere/inst/filters.html} for a full description of the SPHERE filters.}. The probability that the source is a background star, evaluated as described in Sec.~\ref{sec:common_pm}, is also listed. 

\begin{figure*}
    \centering
    \includegraphics[width=0.33\textwidth]{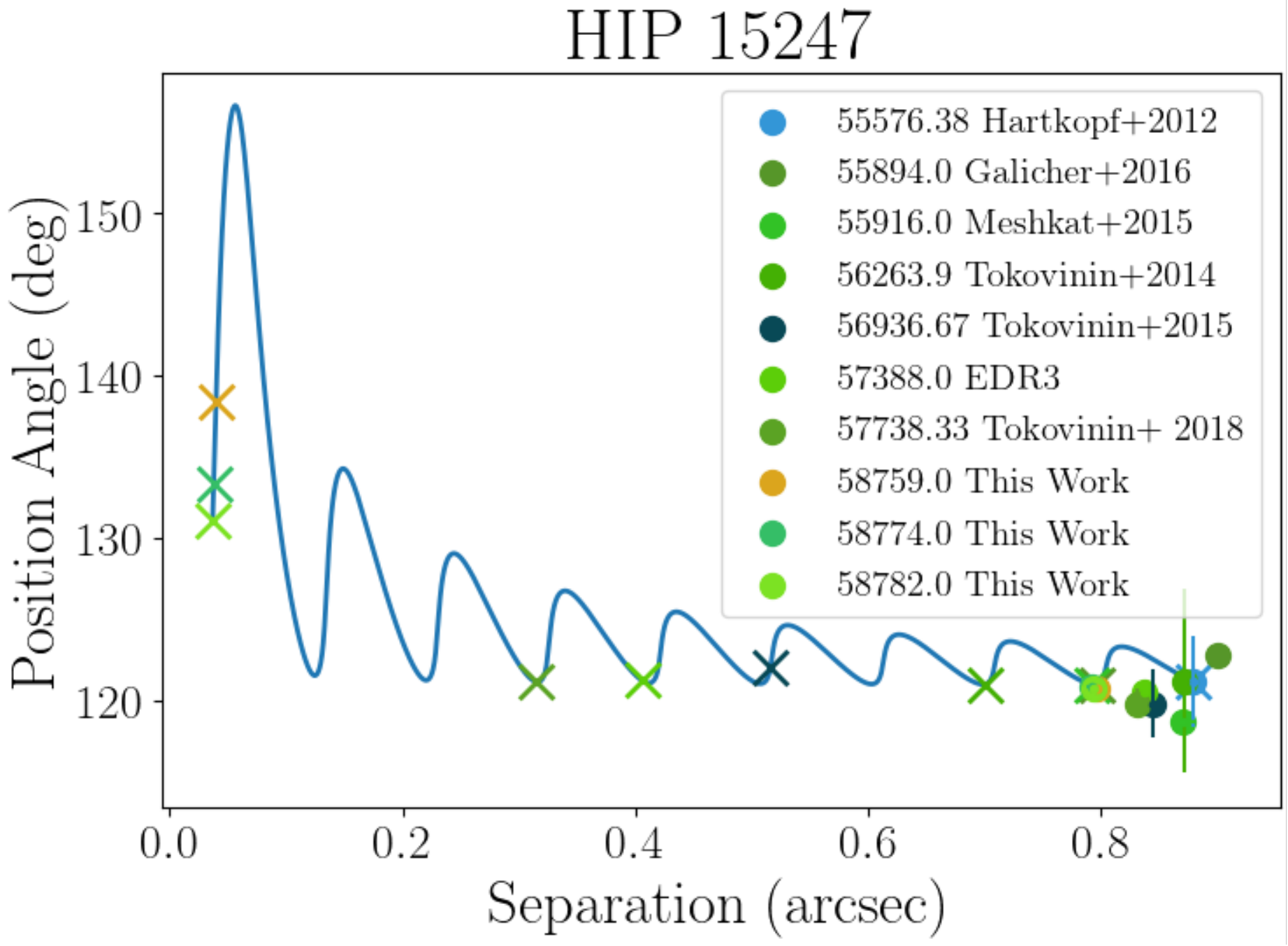}
    \includegraphics[width=0.33\textwidth]{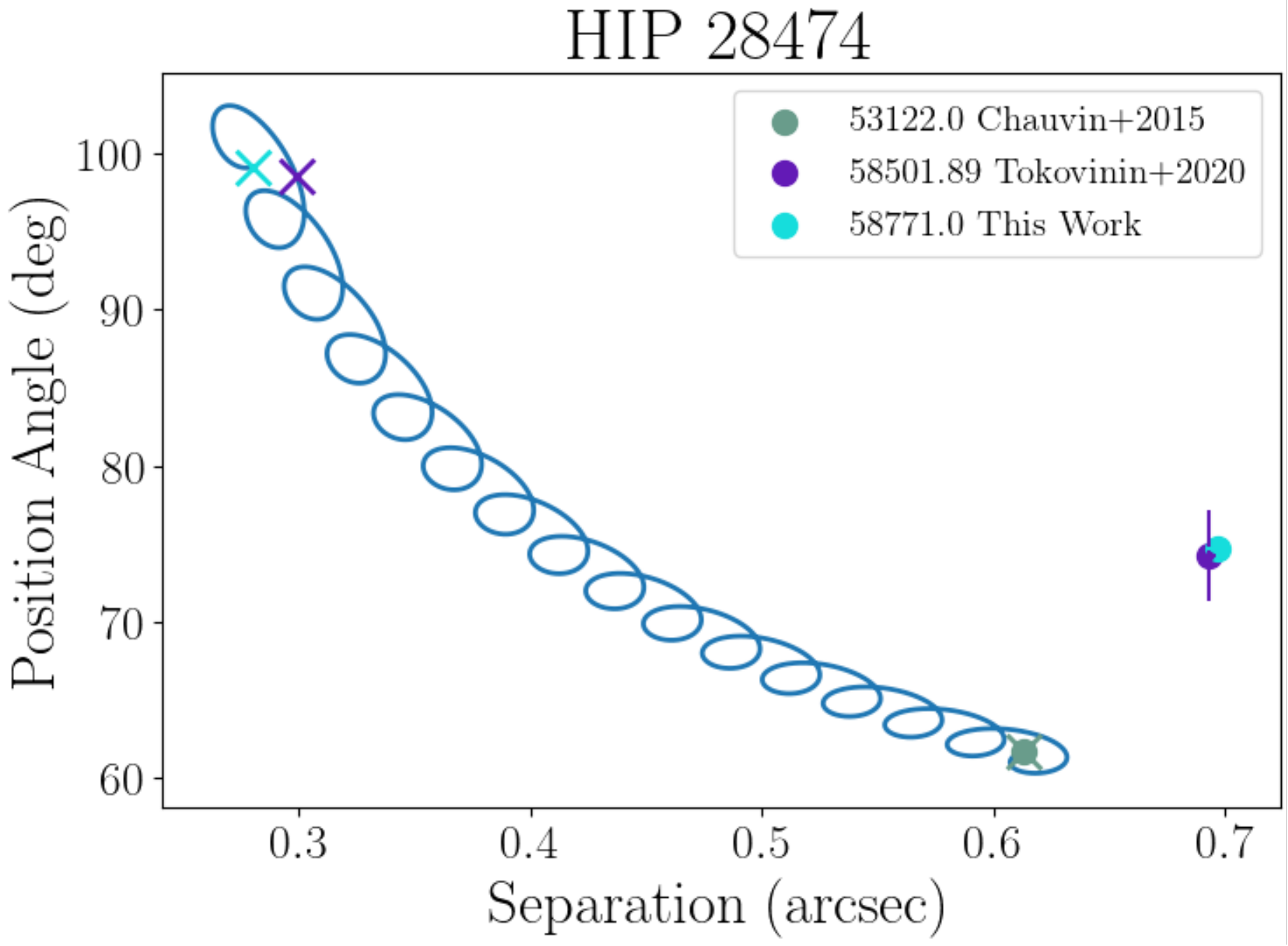}
    \includegraphics[width=0.33\textwidth]{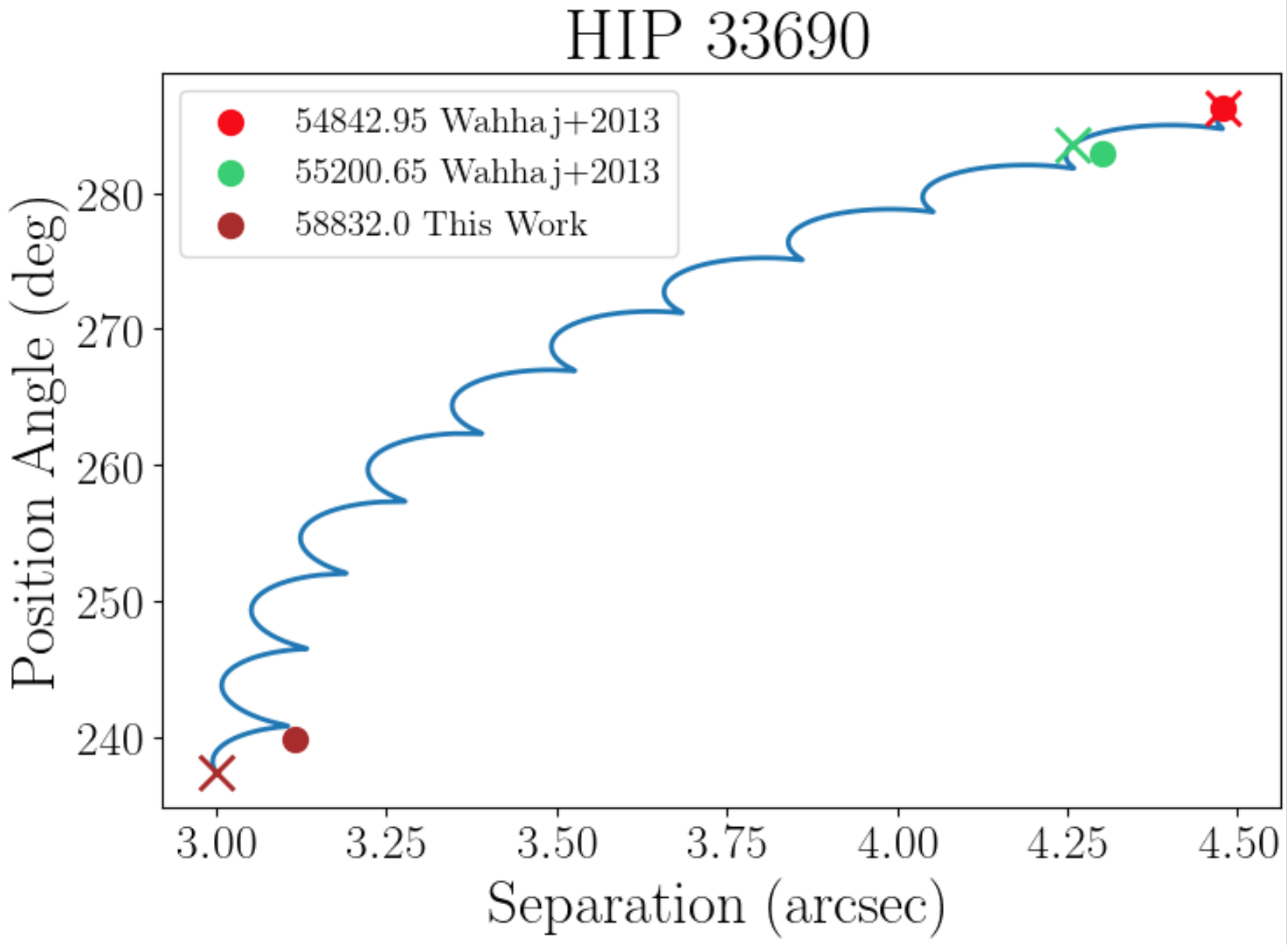}
    \includegraphics[width=0.33\textwidth]{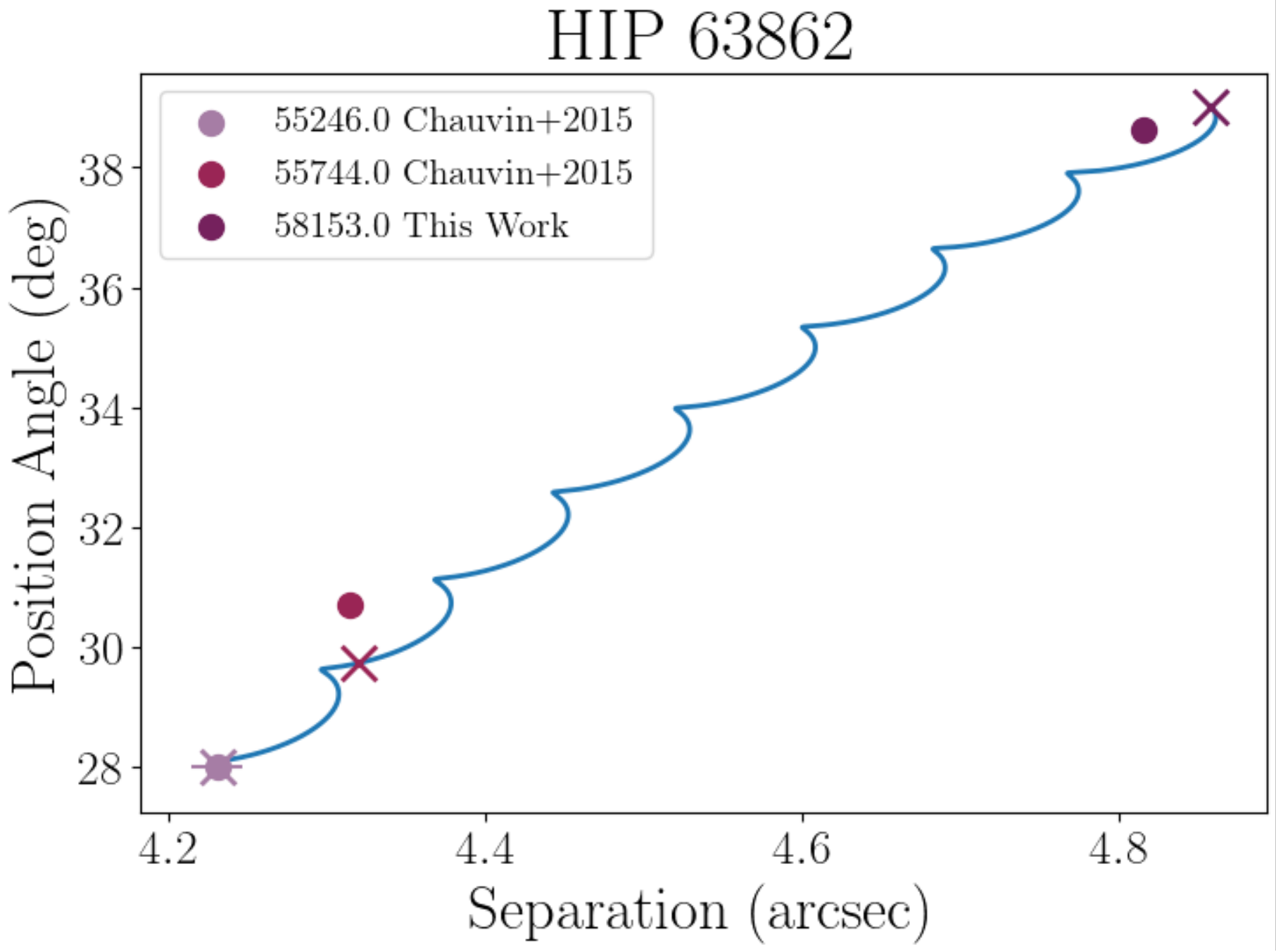}
    \includegraphics[width=0.33\textwidth]{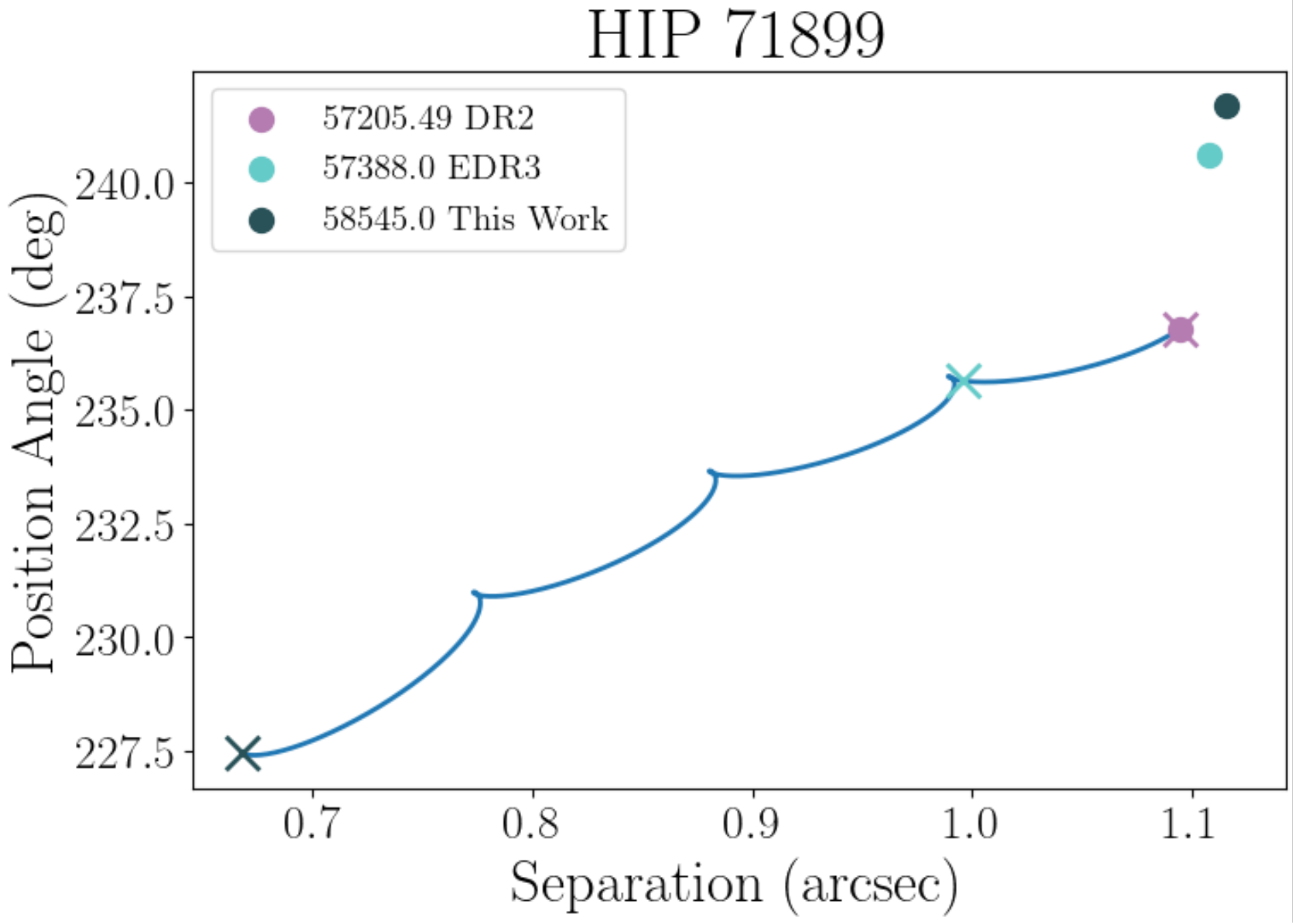}
    \includegraphics[width=0.33\textwidth]{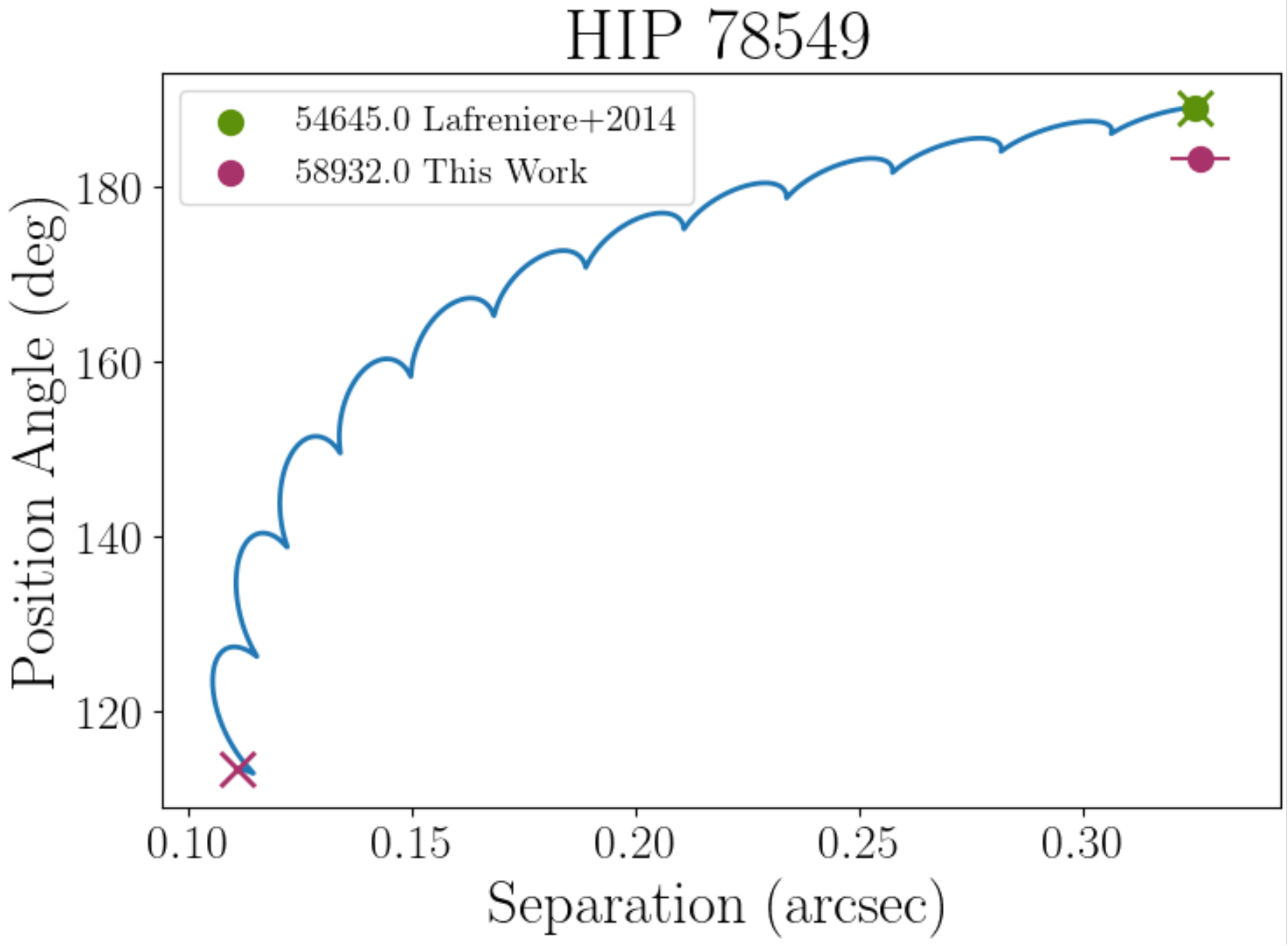}
    \includegraphics[width=0.33\textwidth]{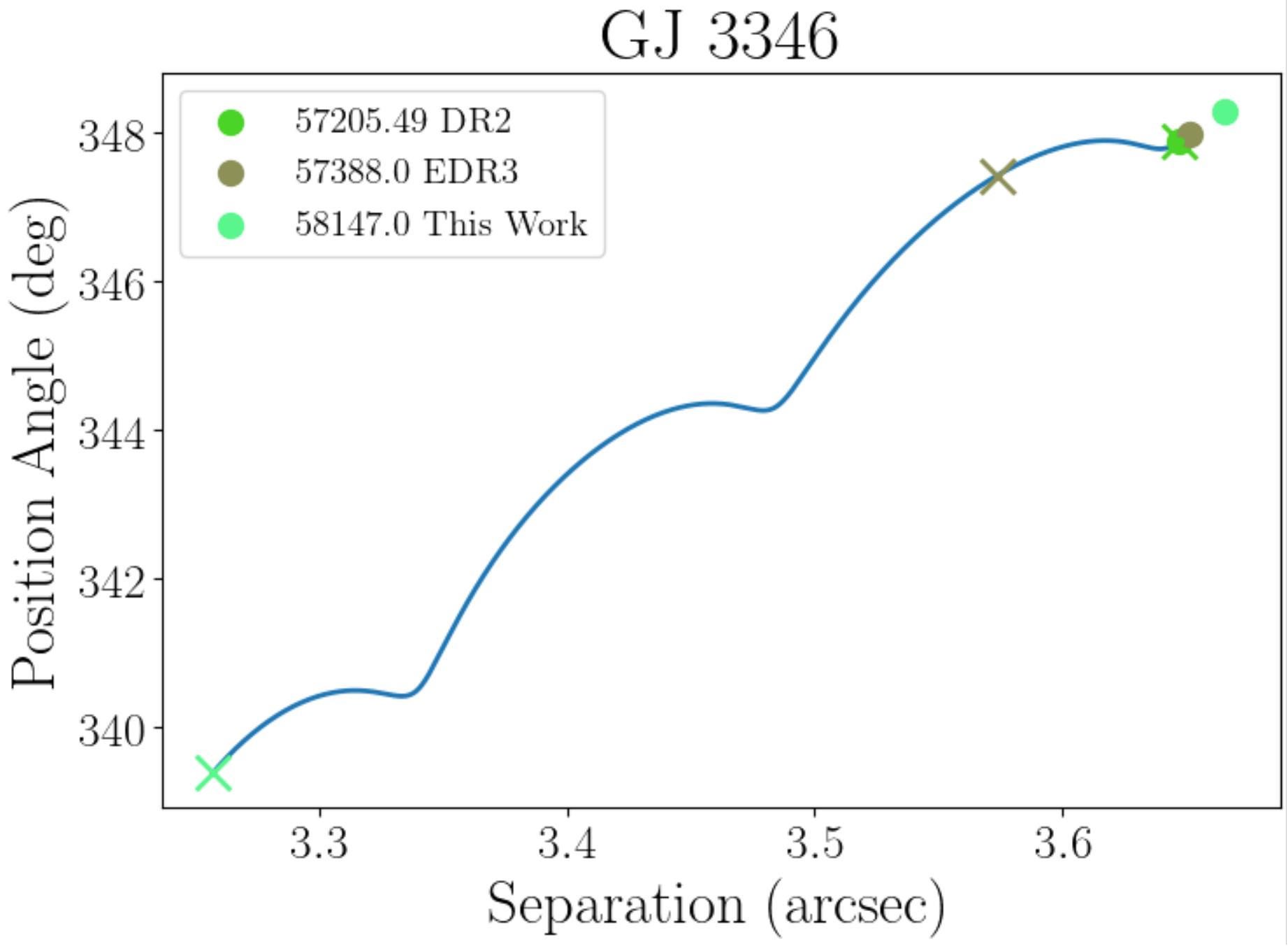}
    \includegraphics[width=0.33\textwidth]{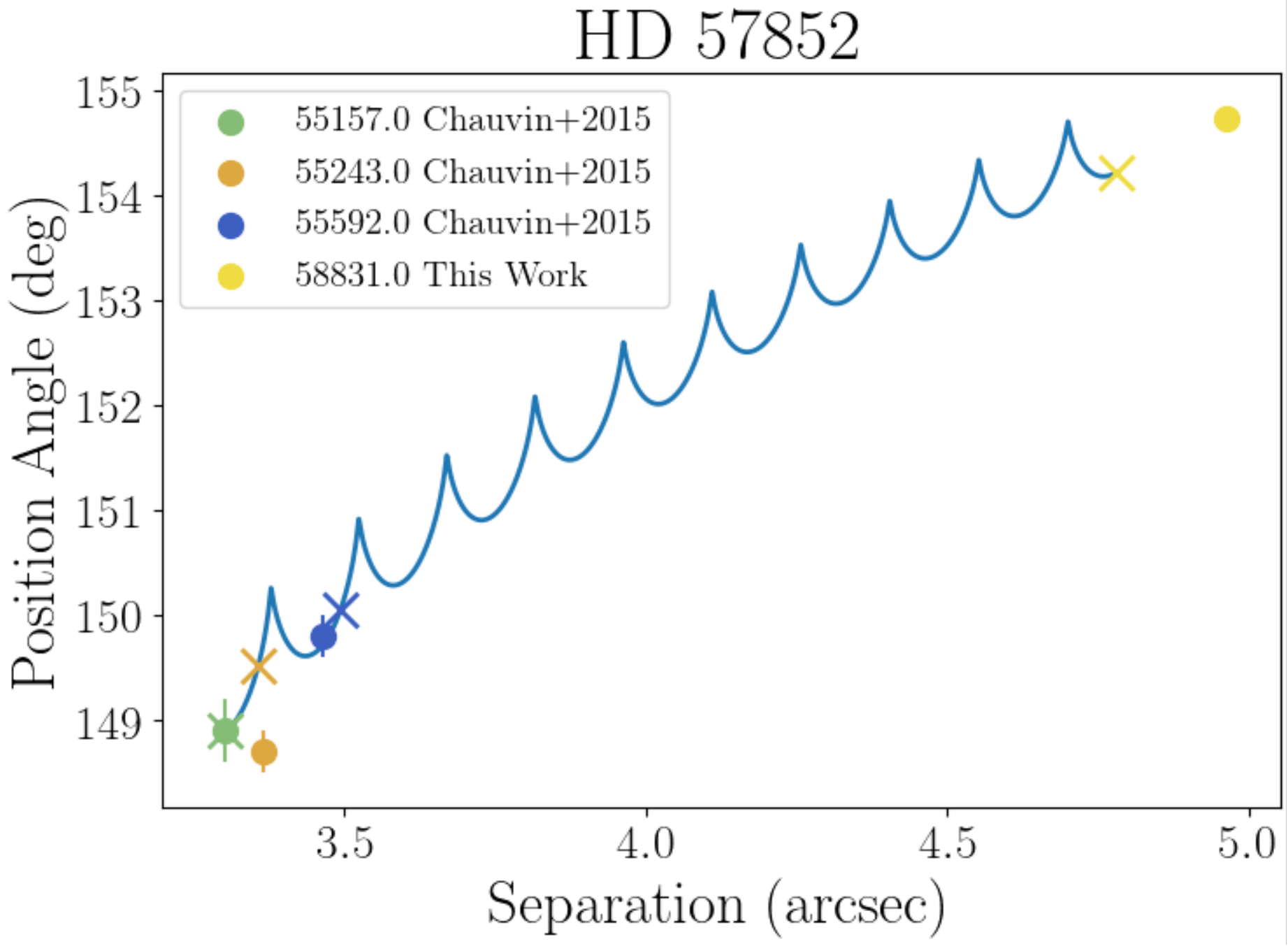}
    \includegraphics[width=0.33\textwidth]{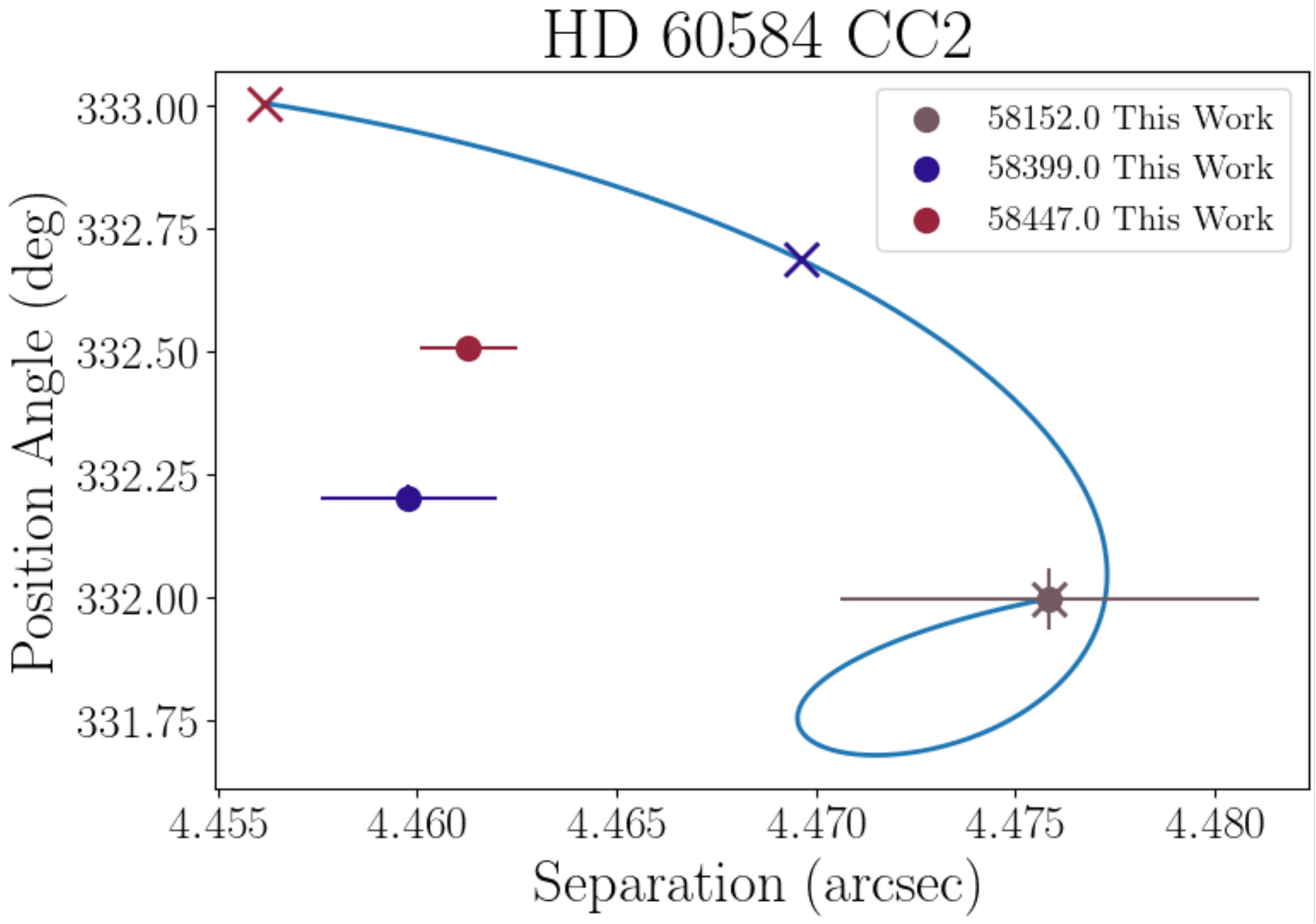}
    \caption{Common proper motion analysis of all our candidate companions with multiple epochs (reported in Table~\ref{tab:multi_epoch}). 
    The blue line shows the motion of a background object relative to the target, based on the EDR3 parallax and proper motion of the primary over the same time frame.
    In all panels, the filled circles show the measured separation and position angle of the companions at each epoch, colour coded as explained in the legends. The crosses indicates the expected position of a background object at the same epoch, following the same colour code.
    The companions for HIP~28474, HIP~78549, HIP~15274, HIP~71899 and GJ~3346 (updated from the dedicated work published in \citet{bonavita2020b}, which only included {\it Gaia}DR2 epoch), 
    are clearly found to be co-moving with our targets.
    }
    \label{fig:CPM_bound}
\end{figure*}

\subsection{Common proper motion confirmation}
\label{sec:common_pm}
Multiple epochs, used to clarify the bound or background nature of our candidates, were available for 9 of the program stars. Except for HIP~15247 and HD~60584, which were re-observed with SPHERE as part of the program, all additional epochs were retrieved from other surveys, catalogues (including Gaia) or papers dedicated to specific objects. The complete list of astrometric measurements for all our systems is presented in Table~\ref{tab:multi_epoch}, together with the references used for each entry. 
Figure~\ref{fig:CPM_bound} shows the resulting common proper motion analyses for both the co-moving and background interlopers. 
Note that for HD~60584 we detected two sources, one at 0.5 arcsecs (CC1, discussed in Sec.~\ref{sec:tyc6539-HIP6734}) and one at $\sim$3 arcsecs (CC2), shown in Fig.~\ref{fig:CPM_bound} and confirmed to be background.
The remaining candidates are bright companions at very small separation and are then very likely physically related as the probability of having such bright background stars at these separations is very low. To confirm this, we used the code described in Section 5.2 of \citet{Chauvin2015} and adapted it to our results to estimate the probability of finding a background contaminant at the given separation and contrast as a function of galactic coordinates by comparison with the prediction of the Besan\c{c}on galactic model \citep{Robin2012}. As expected all the resulting probabilities (reported in the last column of Table~\ref{tab:photoastro}) are below 10$^{-4}$. A further confirmation of common proper motion is also provided by the agreement between the properties of the imaged companions with observed $\Delta \mu$, as discussed in Sec.~\ref{sec:forecast}.

\begin{figure*}
    \centering
    \includegraphics[width=0.45\textwidth]{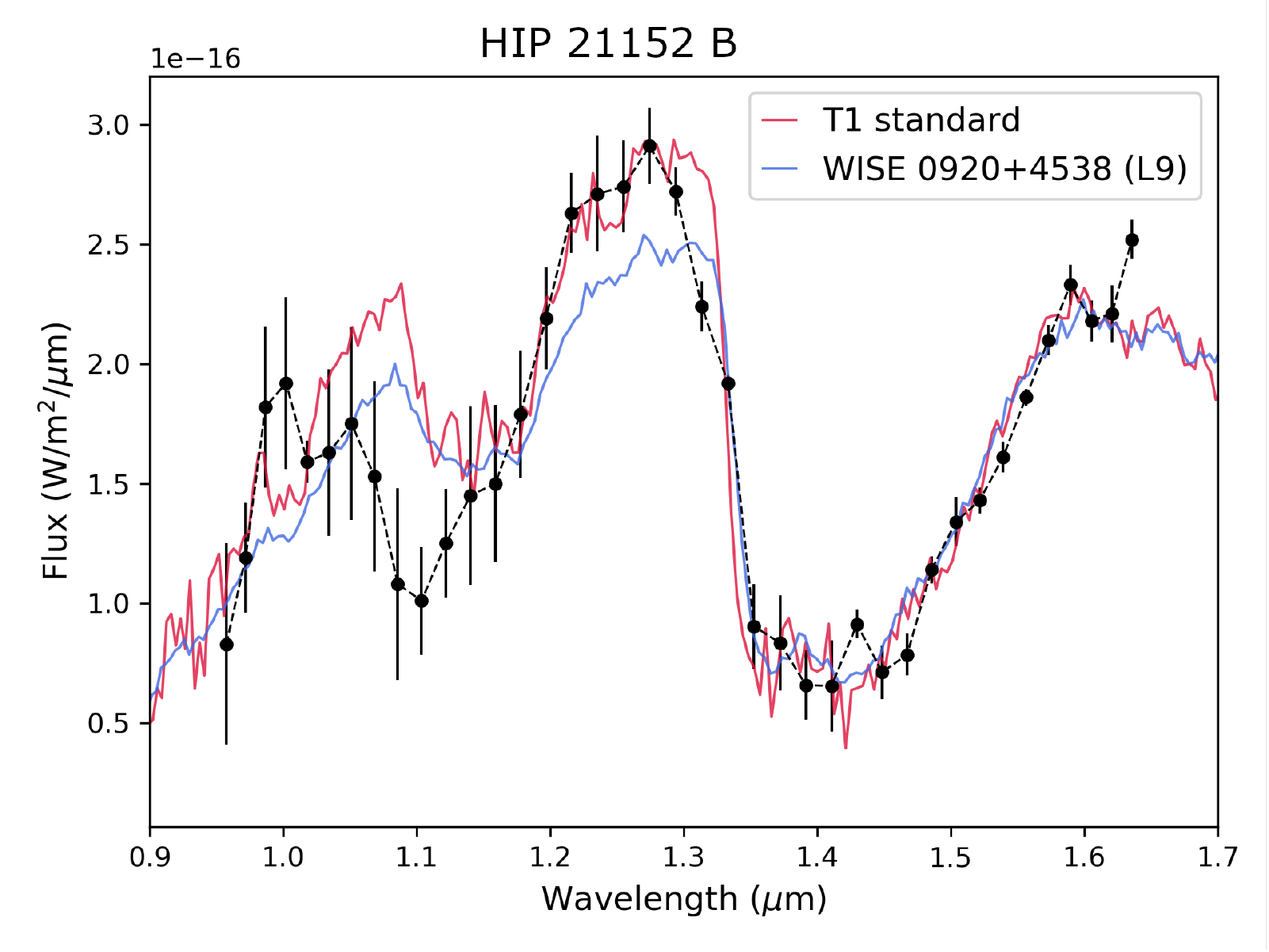}
    \includegraphics[width=0.45\textwidth]{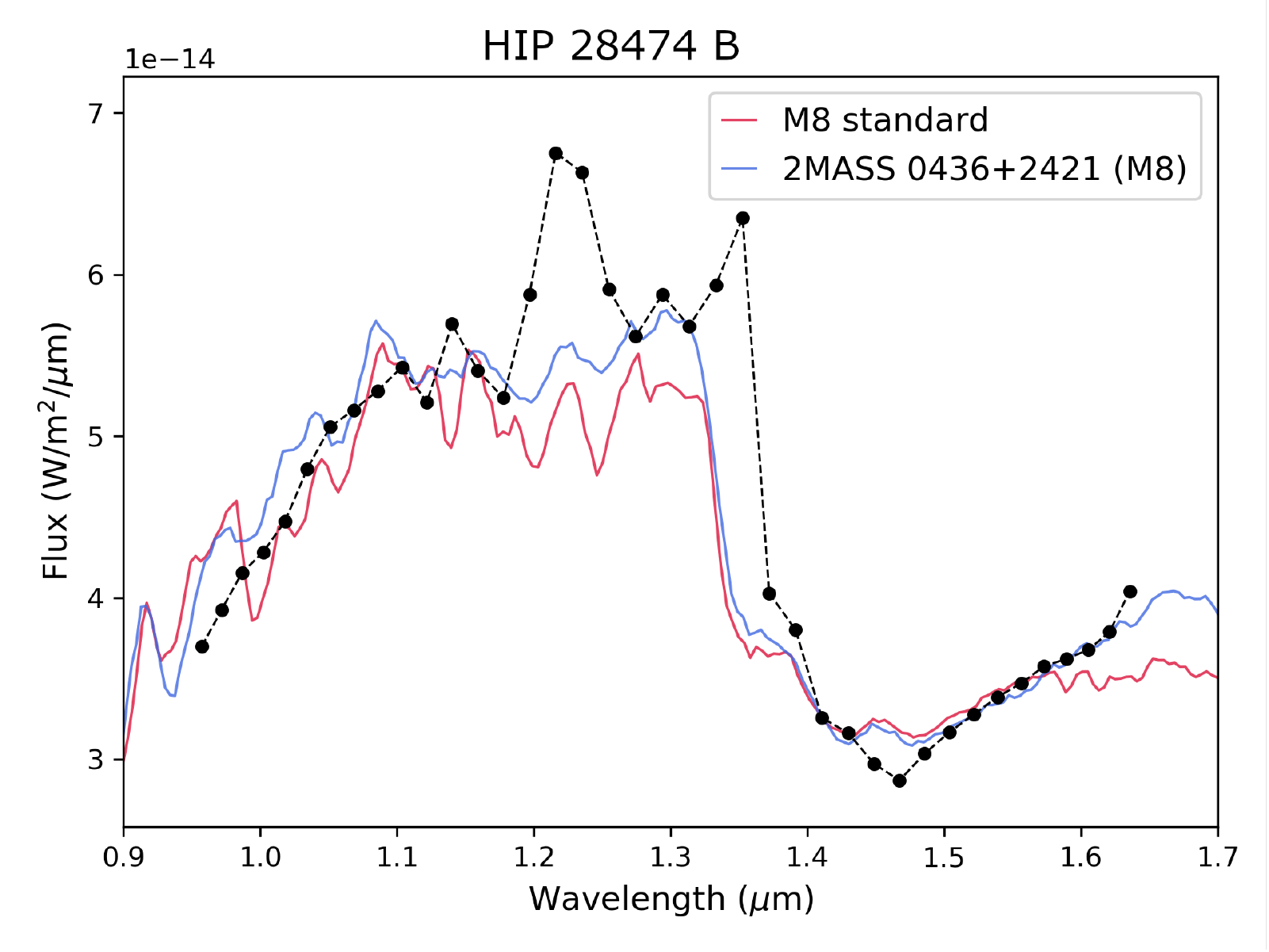}
    \includegraphics[width=0.45\textwidth]{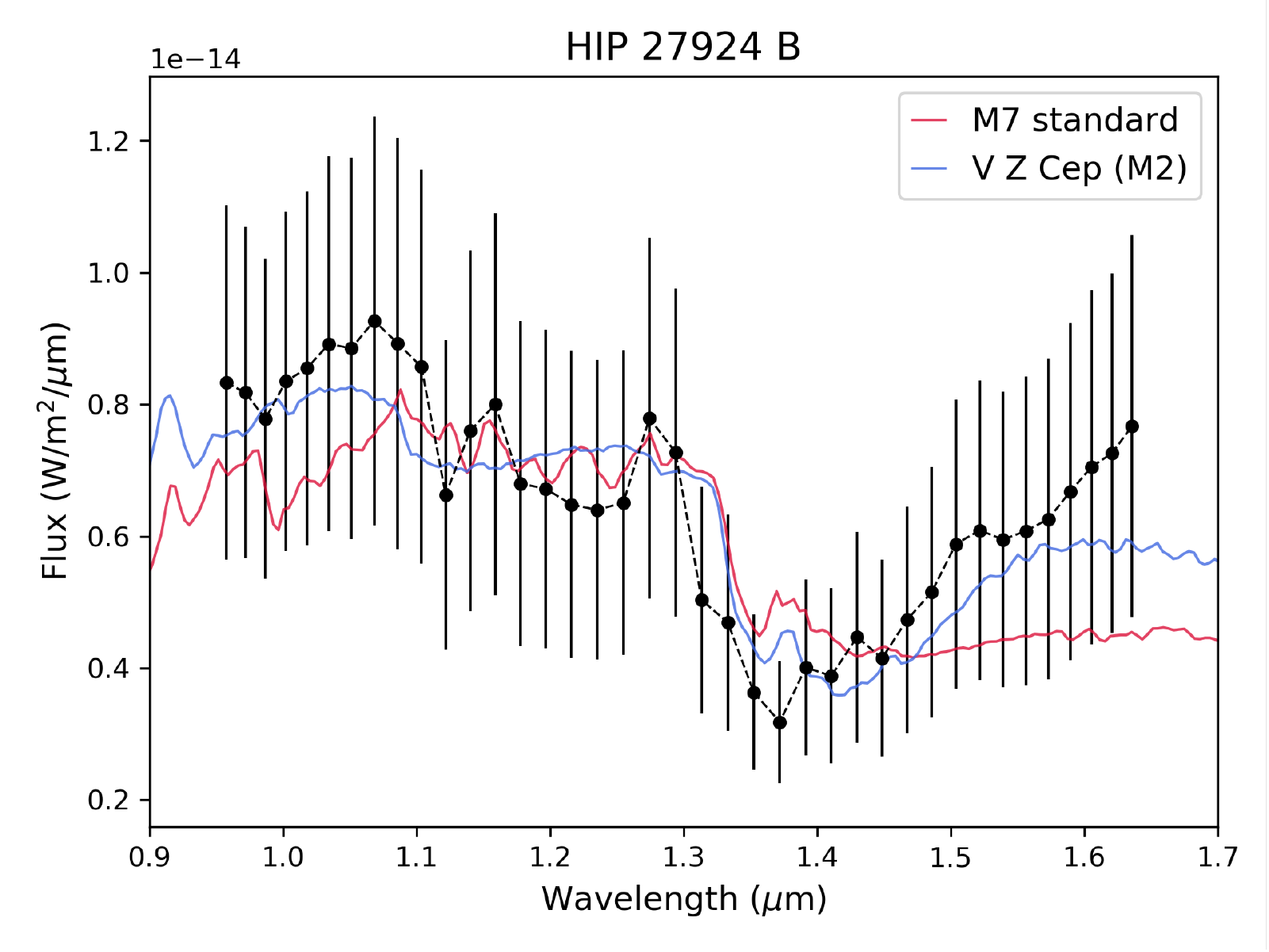}
    \includegraphics[width=0.45\textwidth]{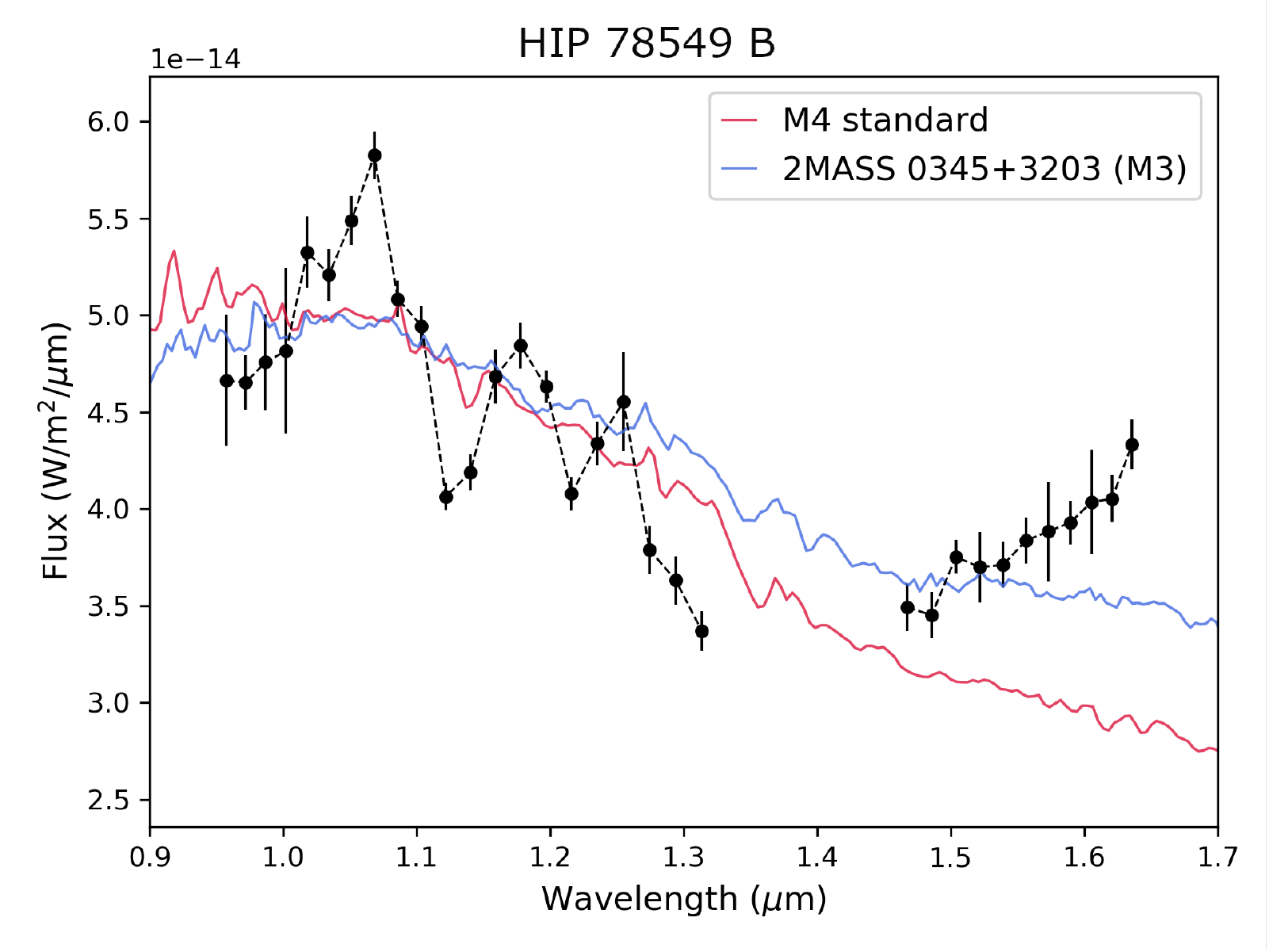}
    \includegraphics[width=0.45\textwidth]{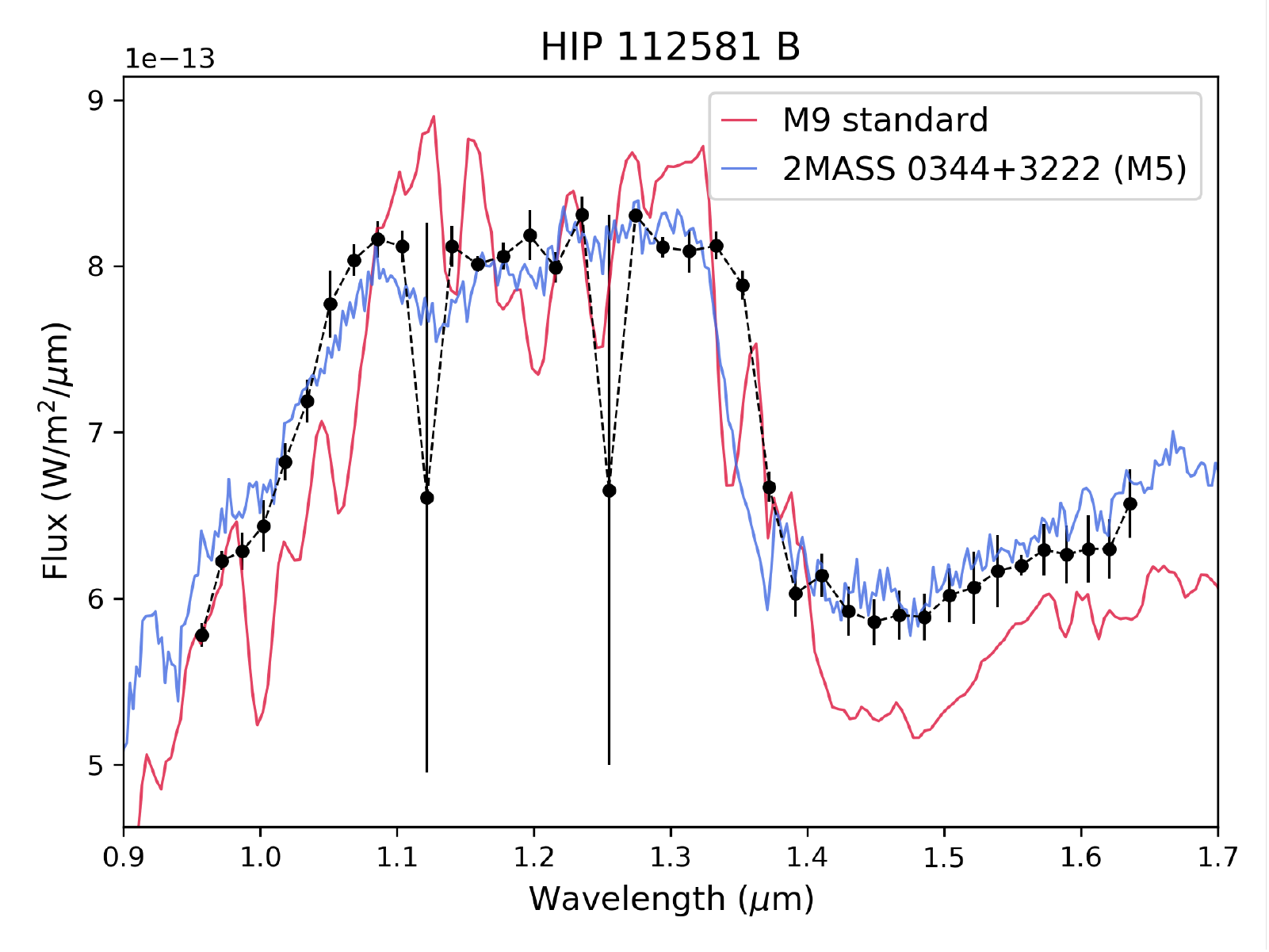}
    \caption{Comparison between the spectra of five of the companions detected at high SNR (black bullets with error bars) with the best fit from SPLAT \citep{SPLAT}, showing in red the best fit among the spectral standards provided in SPLAT, and in blue the overall best fit among all available templates.}
    \label{fig:spectra}
\end{figure*}

\subsection{Companion spectra}\label{sec:spectra}
Spectra for the bright companions (providing relatively high SNR data) were obtained using the IFS data. We used the SpeX Prism Library Analysis Toolkit (SPLAT; \citealp{SPLAT}) to estimate the spectral classification of each companion. We used the built-in spectral fitting function in SPLAT to compare, by minimising the $\chi^2$ value, the observed spectra to both the SPLAT near-infrared spectral standards and to the full library of templates available in SPLAT. 
For bright companions, we simply used the spectrum obtained by rotating and summing the images. For HIP~21152~B, which was not detectable without removing the speckle pattern, we injected negative point spread functions (derived from the flux calibration) on the individual monochromatic images at the average companion position, and changed its intensity minimising the root mean square of the residuals in area of $9\times 9$ pixels centred on this mean position. Figure~\ref{fig:spectra} shows the spectral standard (red) and template (blue) providing the lowest $\chi^2$ values to the observed spectrum of each detected companion. 
Note that the images of HIP~15247~B was saturated at most wavelengths, so it was not possible to obtain an usable spectrum. Among the five remaining companions, four have an estimated mid- to late-M spectral type, while HIP~21152~B is clearly a substellar object and compatible with a late-L or early-T spectrum.\\
The strong absorption feature visible in the spectrum of HIP~78549~B is most likely spurious and the result of the extremely low quality of the wavelength calibration available for this object, particularly between 1.32 and 1.45 $\mu$m. We therefore masked this region of the spectrum while performing the fit. The resulting mid M spectral type is in agreement with the properties estimated from the photometry. 

\begin{figure*}
    \centering
    \includegraphics[width=\textwidth]{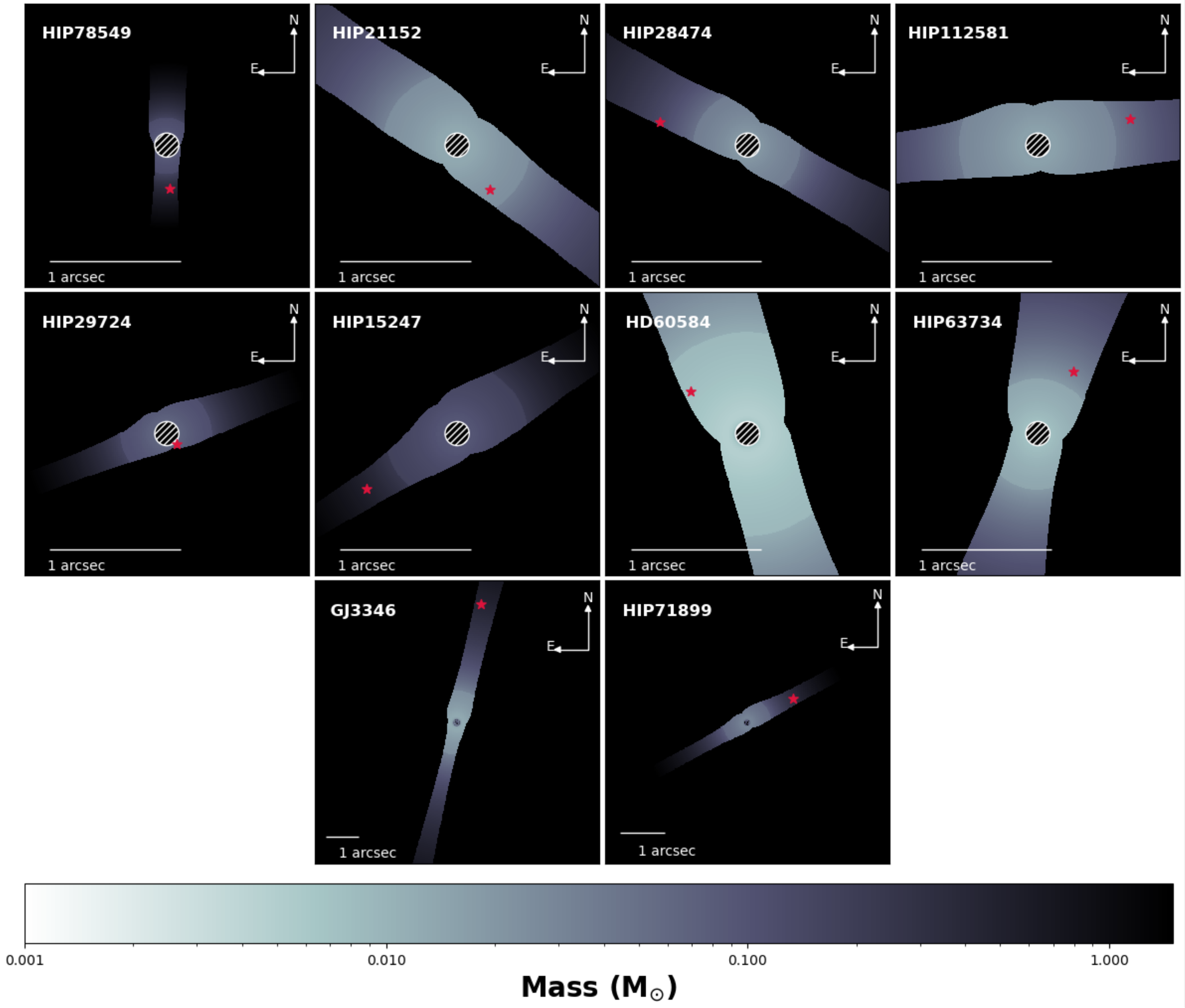}
    \caption{2D maps representing the sky area compatible with the TGAS-EDR3 $\Delta\mu$  reported in Table~\ref{tab:deltamu}. The position of the companions is marked with a red star. Colours are according to the dynamical mass responsible for the $\Delta\mu$ at a give distance; the same logarithmic scale was used for all stars, according to the colour scale shown on the bottom of the figure. The empty area at center is the area covered by the coronagraphic mask.}
    \label{fig:pma_maps}
\end{figure*}

\begin{figure*}
    \centering
    \includegraphics[width=0.45\textwidth]{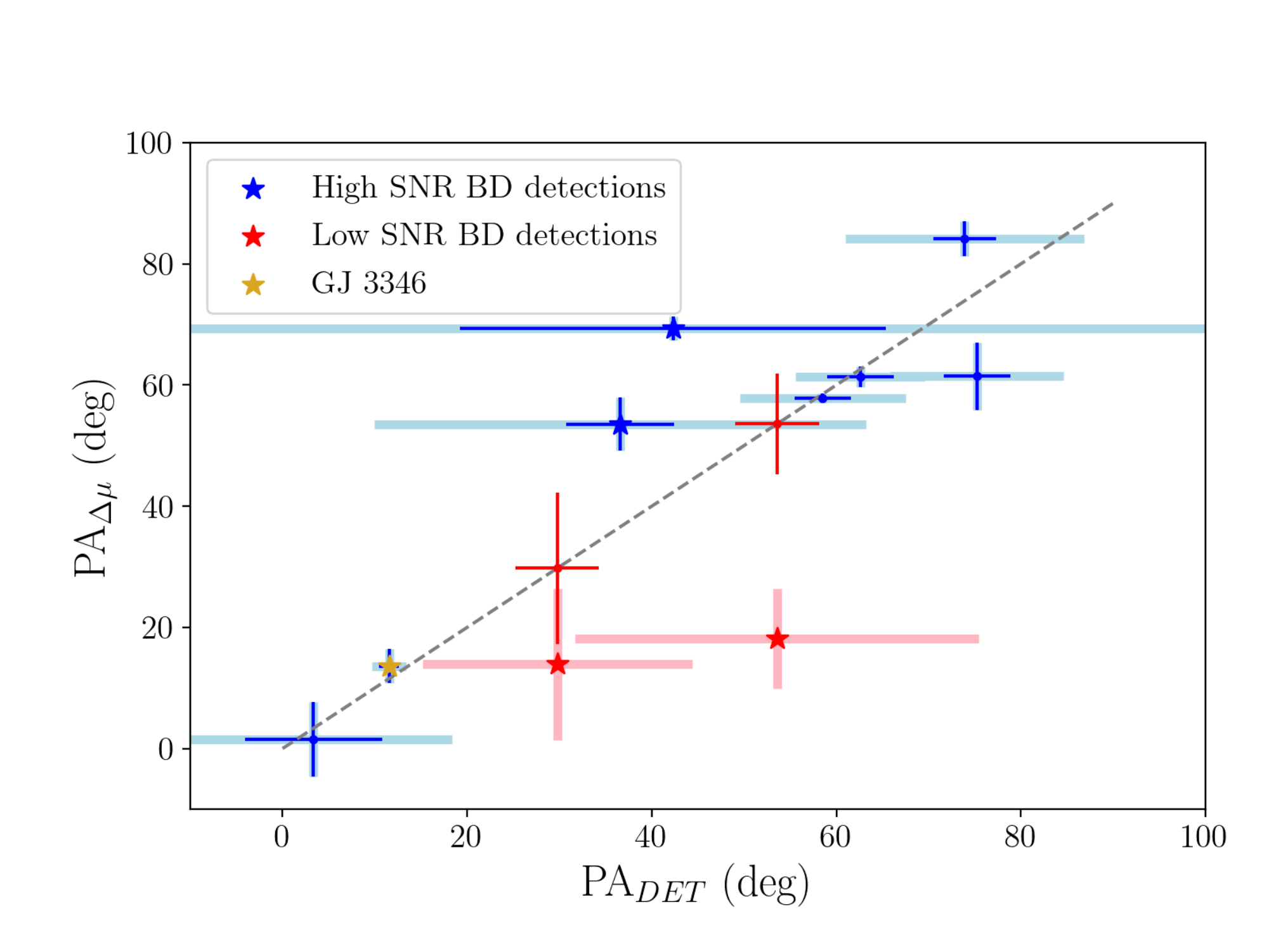}
    \includegraphics[width=0.45\textwidth]{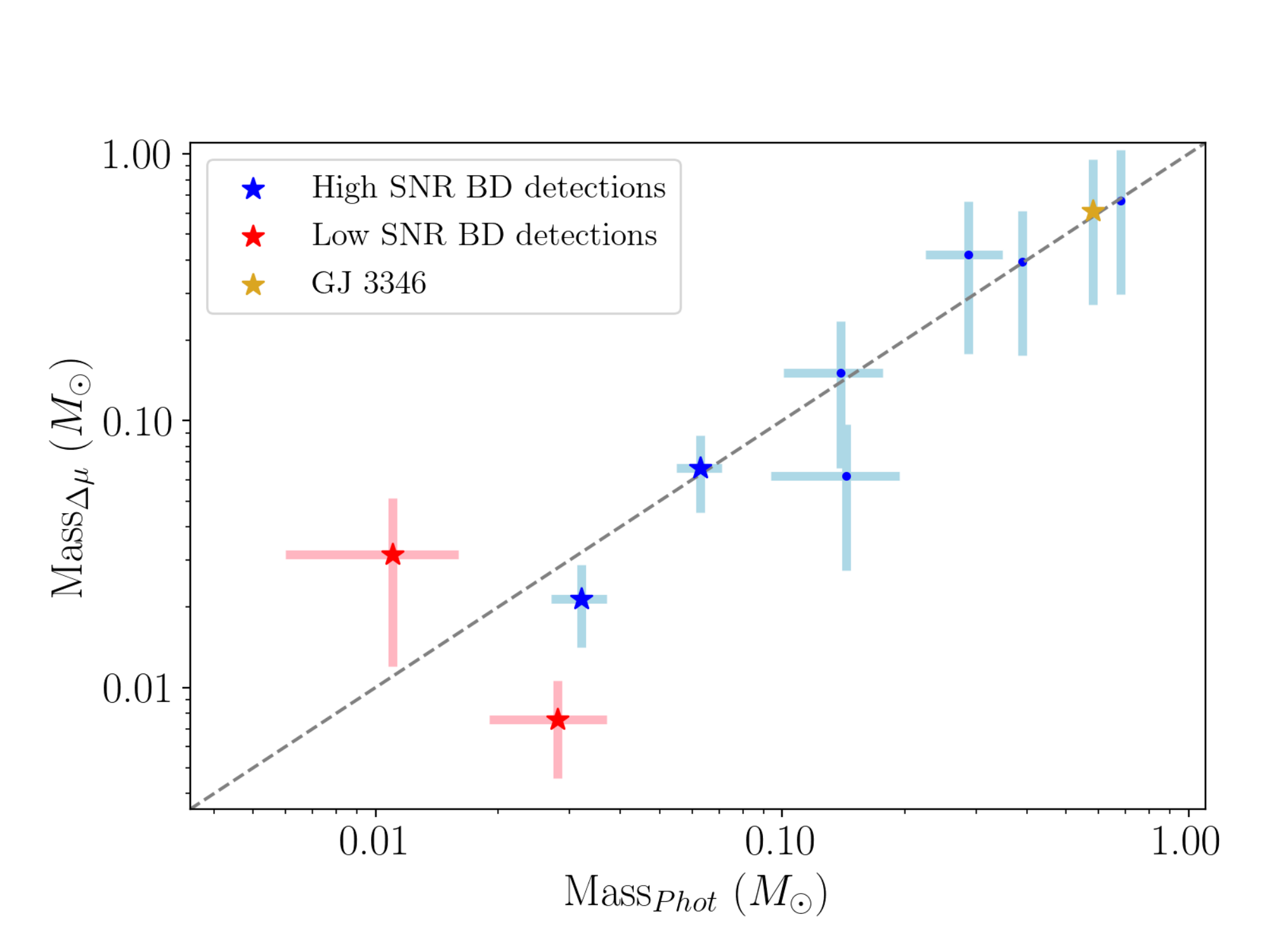}
    \caption{{\bf Left Panel:} PA of the detected companions VS the direction of the $\Delta\mu$. The shaded error bars show the extend of the expected change in angle due to the orbital motion between the SPHERE observation and the {\it Gaia} EDR3 epoch. 
    {\bf Right Panel:} Comparison between the values of the mass of the detected companions derived using FORECAST, and those obtained from photometry and evolutionary models (as described in Sec.~\ref{sec:mstar}). Both values of the masses are reported in Table~\ref{tab:sys_char} for all companions.  In both panels blue symbols are high SNR detections; red symbols are low SNR candidates that needs further confirmation. }
    \label{fig:dangle}
\end{figure*}

\subsection{FORECAST (Finely Optimised REtrieval of Companions of Accelerating STars)}
\label{sec:forecast}
Since the $\Delta \mu$ can be considered as an approximation of the instantaneous acceleration due to an unseen companion, it can be represented as a vector in the plane of the sky directed toward the position of the companion - at the epoch of the latest astrometry observation. 
The position angle of an imaged companion compatible with the $\Delta \mu$ should then be along the same direction of the acceleration, plus or minus the change in angle due to the orbital motion of the companion between the latest astrometry epoch and the imaging one. 
Based on these considerations, it is possible to highlight a region on the plane of the sky where the companion compatible with the $\Delta \mu$ should lie, based on the $\Delta \mu$ orientation, and also associate a value of the companion mass corresponding to each point in the resulting 2D map, based on the $\Delta \mu$ absolute value. 
An example of the $\Delta \mu$ maps obtained with this method using the $\Delta \mu$ at the epoch of $Gaia$ EDR3 (2016.0) is shown in Fig.~\ref{fig:pma_maps}, while a complete description of the method, and of the FORECAST (Finely Optimised REtrieval of Companions of Accelerating STars) code used to produce the maps will be the subject of a dedicated companion publication (Bonavita et al. 2022, in preparation). 
In each image, the position of the detected companion is marked as a star. 
As all companions lie within the allowed region (as shown in the left panel of Fig.~\ref{fig:dangle}, which compares the PAs of the companions and those of the $\Delta \mu$ vectors), we could also obtain a first estimate the expected companion mass at that position and compare it with the values of the value of the mass derived from the photometry. 
The dynamical masses ($M_{\Delta\mu}$) were estimated using the method described in \cite{Kervella2019}, which allows to also take into account the effect of the unknown orbital eccentricity and inclination, as well as that of the observing window smearing, which are instead not taken into account by the COPAINS tool. The comparison with the photometric masses ($M_{Phot}$) is shown in the right panel of Fig.~\ref{fig:dangle}, with the error bars on the values of $M_{\Delta\mu}$ taking into account the errors on the $\Delta\mu$ and companion position, and those on $M_{Phot}$ mainly arising from the uncertainties on the stellar ages.
The good agreement between the values shows the potential of our method in providing a high number of potential new benchmark objects. 

\subsection{HIP 21152~B: a new bound substellar companion in the Hyades}
\label{sec:hd28736}
The companion to HIP~21152 is clearly the most interesting of our high SNR detections.
The IRDIS photometry and the IFS spectra, shown in Fig.~\ref{fig:spectra}, point towards an early T spectral type. The mass of the object, as judged from its spectral type and age, and adopting the models from \citet{baraffe2015}\footnote{New atmospheric models have been recently developed by \cite{Phillips:2020AA}. However, such models (known as ATMO2020) are mostly valid for relatively old ultra-cool objects (late-T and Y companions). Therefore, given the age and brightness of HIP~21152~B as well as the estimated spectral type, we decided that the models from \citet{baraffe2015} would be more suited for the mass estimate in this case. The same applies to the other BD companions described in Sec.~\ref{sec:hip29724} and \ref{sec:tyc6539-HIP6734}}, is estimated to be $0.032\pm0.005~M_{\odot}$. This is also in good agreement with the mass derived from the FORECAST analysis, which is $0.021\pm0.007~M_{\odot}$. \\
The Lithium-depletion boundary, an observational limit that separates low-mass stars from brown dwarfs based on their ability to burn Li in their cores, has been estimated to occur around spectral type L3.5 - L4 in Hyades \citep{martin2018}. Currently, about a dozen objects with spectral type L3.5 or later claimed as members of the cluster. \citet{lodieu2014} published spectra of 12 L-dwarf candidates from \citet{hogan2008}, and confirmed one of them as a potential brown dwarf member of Hyades, with spectral type L3.5. The same objects has later been listed as a L4 member of Hyades, with Li absorption detected in the spectra by \citet{martin2018}. Four more L-type members (two L5 and two L6) have been confirmed in \citet{schneider2017,perez-garrido2017,perez-garrido2018}. 
\citet{bouvier2008} were the first to report the existence of T-type candidate members in Hyades. The membership of one of them (CFHT-Hy-21) has been confirmed by \citet{lodieu2014}, whereas the other one (CFHT-Hy-20) is still considered a candidate \citep{lodieu2014,zhang2021}, given  the lack of radial velocity measurement. Four T-type probable members have been reported in \citet{zhang2021}, spanning the spectral type range T2 to T6.5. The multiplicity of Hyades stars has been assessed through imaging \citep{reid&gizis1997,patience1998,siegler2003,duchene2013b}, high-resolution spectroscopy \citep{reid&mahoney2000}, a combination of the two \citep{guenther05}, and $Gaia$ astrometry \citep{deacon20}. While several binary systems consisting of two  substellar objects, or objects close to the (sub)stellar border, were reported in \citet{siegler2003,reid&mahoney2000,duchene2013b}, no substellar companions to FGK stars have so far been reported in Hyades, making HIP~21152~B the first of its kind. HIP~22152~B thus occupies a unique place in the luminosity-age parameter space compared to the known population of directly-imaged systems, making it a highly valuable benchmark for empirical constraints to theoretical models.\\
BD companions such HIP~22152~B have been shown in the past to be themselves close pairs of substellar objects
\citep[see e.g.][]{Martin:2000ApJL, Potter:2002ApJL} and optimised PSF subtraction methods have been recently developed to highlight the presence of close companions to BDs detected with SPHERE \citep[see e.g.][]{Lazzoni:2020}. 
Such methods rely on the detection and analysis of features such as elongation in the companion's PSF incompatible with the expected effects of ADI. 
Although a detailed analysis would benefit from a new set of higher quality data, we can say that no such features were detected in our images for HIP~21152~B. 
Moreover, to be on stable orbits, possible additional companions would need to have separations within HIP~21152~B's Hill Radius, which we estimated to be roughly at 3.57 au, for the optimistic case of a circular orbit. At the distance of HIP~21152~B ($\sim$43 pc), this corresponds to a projected separation of roughly 82 mas. While this target has good potential for the detection of additional (possibly less massive) companions, an analysis like the one described by \cite{Lazzoni:2020} will require higher quality data than those presented in this work.\\

\subsection{HIP~29724~B: a new high mass brown dwarf}
\label{sec:hip29724}
We also detected a very close (99.9$\pm$1.5~mas) companion to HIP~29724. 
The photometry suggests a mass of 0.063$\pm$0.008~M$_{\odot}$, thus placing HIP~29724~B also in the sub-stellar regime. 
The FORECAST analysis confirms that the companion is compatible with the observed TGAS-EDR3 $\Delta\mu$, both in terms of position and corresponding mass ($0.067\pm0.021~M_{\odot}$). With a projected separation of just 6.3 au, RV monitoring could significantly contribute to the refinement of the BD orbit and dynamical mass.
However, there are no high-precision RV measurements available up to now.
Sparse RV data over several decades \citep{nordstrom2004,sacy,GaiaDR2}
are constant within 1.2 km/s.
The spectrum of the companion, shown in Fig.~\ref{fig:spectra}, is compatible with a late M spectral type, although a later spectral type is not completely excluded, given the uncertainties caused by the companion's proximity to the edge of the coronagraph.\\
 As it was the case for HIP~21152~B, the PSF of HIP~29724~B does not show any elongation beyond what expected as result of the ADI. We estimated for this companion an Hill Radius of 1.71~au (27.2 mas, assuming a circular orbit) thus limiting the detectability of possible additional companions to roughly equal-mass ones, considering the spatial resolution of our images.

\subsection{HD~60584~B and HIP~63734~B: two more potential low mass brown dwarf candidate companions}
\label{sec:tyc6539-HIP6734}
The $\Delta \mu$ maps obtained with FORECAST are not only useful to confirm the nature of the candidates found in objects with highly significant $\Delta\mu$, but could in principle also be used as \emph{finding charts} to highlight and retrieve possible additional companions appearing in the imaging data at a SNR lower than the threshold required for a confirmed detection, which would otherwise be overlooked. 
We had in fact noticed that all our high SNR detections were around stars for which the ratio between the $\Delta\mu$ absolute value and its error (hereafter $SNR_{\Delta\mu}$) is higher than 10. So we decided to produce the FORECAST maps also for the targets with low or intermediate $SNR_\mathrm{Max}$ (HD~60584, HIP~63734, and HIP~22506) as a test for their possible use as \emph{finding charts}. \\ 
While the data for HIP~63734 were taken in fairly good atmospheric conditions, the target was observed quite far from meridian passage so that field rotation is limited (only 7.13 degrees). The analysis of the IFS data with ASDI-PCA revealed a candidate companion with $SNR\sim 8.5$, at separation $555\pm 2$~mas and PA=$329.6\pm 0.2$~degree, with a contrast of dJ=12.08 and dH=11.26. The comparison with \citet{baraffe2003} models yields an evolutionary mass of $0.011\pm0.005$M$_{\odot}$. While the PA of the object is fully compatible with that of the $\Delta\mu$, the corresponding companion mass is a bit higher ($0.032\pm0.020$M$_{\odot}$) is a bit higher than the evolutionary one. However, this detection is uncertain because the small rotation angle makes the noise distribution quite different from a Gaussian. \\
The case of HD~60584 is slightly more complicated. Although there were four available epochs, the resulting IFS data were all of relatively poor quality. 
The second and third epochs were taken at about one month interval; since we did not expect a large orbital motion between them, we combined them as a single observation. This allowed us to identify a point source compatible with the FORECAST predictions at a separation 543$\pm$5~mas and PA=52.5$\pm0.5$ degrees, with a SNR of 4.9. 
At that position, the mass of a companion compatible with the $\Delta\mu$ would have to be below 0.01~M$_{\odot}$ which, at the relatively high age of the targets would mean that the companion would have to be fairly faint. 
While the IFS photometry points towards a higher mass ($0.028\pm0.009$M$_{\odot}$), given the high uncertainties due to the image and calibration quality and the very field small rotation angle we still deemed it acceptable.
As for the other BDs, no evident elongation pointing towards possible additional companions was observed for the PSF of HIP~63734~B and HD~60584~B. However, given the low SNR of the detections, we estimate that the only potentially detectable companions within the Hill Radius of these BDs (103 and 80 mas, respectively) would have been equal-mass ones.\\
Although further investigation is required to confirm the nature of both HD 60584~B and HIP~63734~B, mostly due to the poor quality of the imaging data, these additional detections clearly show the power of the approach, which pushes the $\Delta\mu$ method towards companions with smaller masses, whose detection is more uncertain and likely to be below the usual 10~$\sigma$ threshold used for automatic retrieval of point sources in imaging data. Although repeated observations will be needed to confirm low SNR imaging detections, with the use of FORECAST a single observation is potentially enough to confirm that a companion observed in high contrast imaging is the one responsible for the $\Delta \mu$, and therefore co-moving. 
Other low SNR point sources were retrieved around HIP~22506, but were discarded mostly based on the strong disagreement between their brightness and the value of the mass compatible with the position within the FORECAST map. This also shows the power of FORECAST in terms of vetting of possible background or spurious sources.

\section{Discussion}    
\label{sec:discussion}
The COPAINS selection method has proven very successful in ensuring a high detection rate in both the stellar and sub-stellar regime.
We detected a total of 14 candidate companions. Two were known binaries (HIP 15247~B and HIP 78549~B) and four were identified as background sources thanks to additional available epochs found in the literature, which also allowed us to confirm the common proper motion nature of four additional new stellar companions, including the white dwarf companion to GJ~3346, described in \cite{bonavita2020b}.
The masses of the remaining four candidates, derived using the available photometry and the evolutionary models from \cite{baraffe2015}, place them in the sub-stellar regime. 
Such high sub-stellar companion detection rate confirms the efficiency of the COPAINS selection method, as well as the new FORECAST code used to confirm their nature, based on the agreement between their properties with the predictions based on the measured $\Delta\mu$. 

\subsection{Comparison with blind surveys}\label{sec:blind_compare}
To quantify the improvement in terms of detection rate compared to blind surveys, we compared our results with those from the first 150 targets from the SHINE survey \citep{Vigan21}. 
Only 93 of the SHINE-150 targets are included in TGAS and DR2, and only 13 have $\Delta\mu$ more significant than 3~$\sigma$, and would have therefore been selected for COPAINS. 
Three of these SHINE $\Delta\mu$ stars have sub-stellar companions, including HIP~65426, one of the two new SHINE detections. 
The resulting sub-stellar companions frequency is then $\sim25\%$ which is significantly higher than what obtained with the blind approach ($\sim9\%$ without any correction due to prior knowledge about the companions, see \cite{Vigan21} for details), once again showing the efficiency of the COPAINS selection method, especially when combined with the FORECAST maps.

\subsection{Limitations} \label{sec:limitations}
We have demonstrated with our campaign the power of using informed target selection processes such as the \texttt{COPAINS} tool \citep{Fontanive2019} to identify new directly-imaged companions. The high detection rate obtained here strongly validates the use of such approaches in survey designs, despite the numerous assumptions and limitations of the work conducted in this pilot survey. Indeed, our original sample selection considered catalogue proper motions taken at face value. Instead, a more accurate approach would require placing all measurements in the same reference frame and at the same epoch, such as the Hipparcos-$Gaia$ Catalog of Accelerations (HGCA) defined by \citet{Brandt2018,Brandt2021}. These catalogues provides Hipparcos and $Gaia$ DR2/EDR3 proper motions, as well as a $Gaia$-Hipparcos scaled positional differences (close to the TGAS proper motions), placing all proper motions at the epochs of {\it Gaia}DR2 and EDR3, respectively, with recalibrated uncertainties. Nonetheless, given the number of approximations made in \texttt{COPAINS}, these differences were found to be negligible, especially given the use made of the resulting computed trends (i.e., visual selection based on comparisons to expected detection limits). As \texttt{COPAINS} considers long-term proper motions as representative of the system's center-of-mass motion, and short-term measurements as the instantaneous reflex motion of the host, the considered $\Delta\mu$ are only good approximations for systems with orbital periods that roughly match the long- and short-term timescales of the considered astrometric catalogues (see \citealp{Fontanive2019}). These limitations are further added to the adopted eccentricity distribution, primarily impacting the width of the computed solutions, and the fact that the approach assumes face-on orbits, which implies that estimated trends actually provide lower mass limits for a given separation. While information from the now available HGCA catalogues \citep{Brandt2018,Brandt2021} or proper motion anomalies measured by \citet{Kervella2019,Kervella2022} would therefore provide more robust and reliable astrometric trends to use for selection purposes (and should consistently be used for orbital and dynamical mass constraints), our results were not impacted by the use of un-corrected catalogue values, and we have nonetheless shown that the idea behind our method offers a highly promising pathway for future observing programs.

\begin{figure*}
    \centering
    \includegraphics[width=0.9\textwidth]{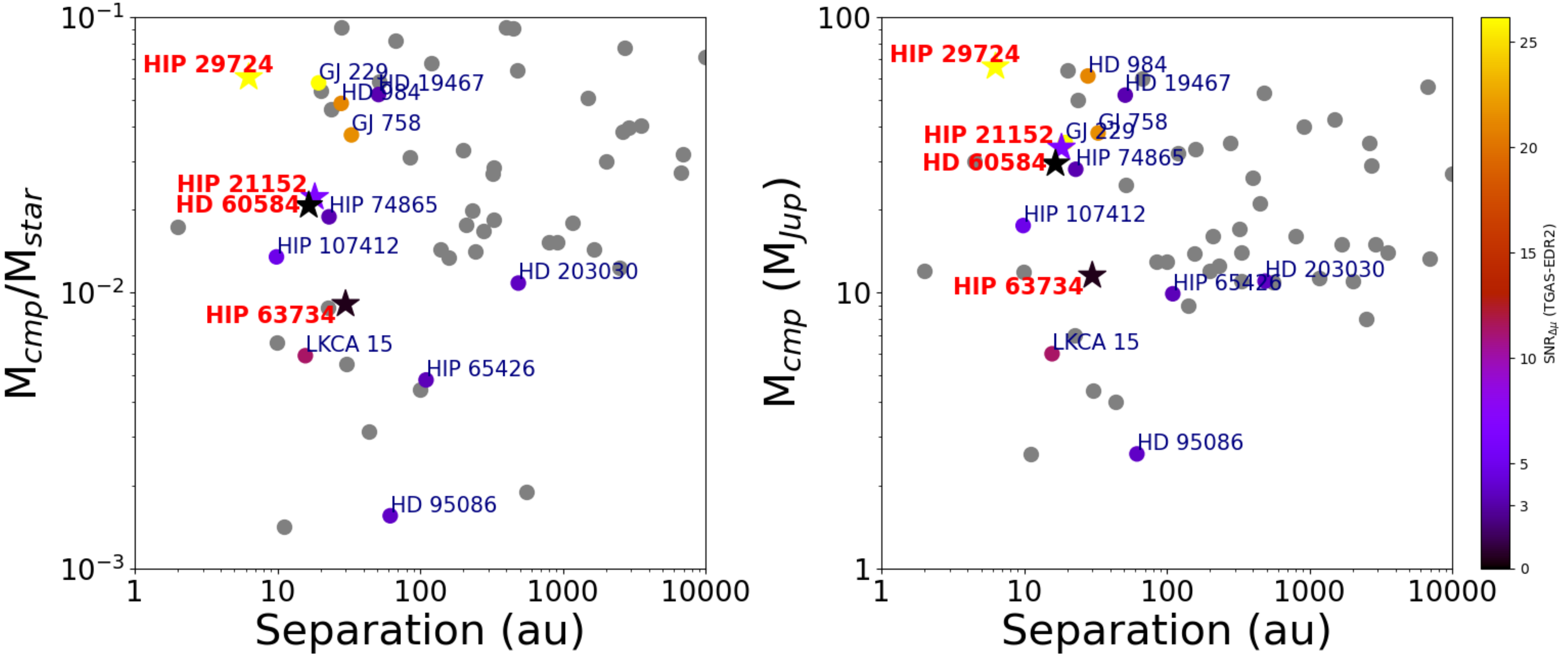}
    \caption{Mass ratio (M$_{cmp}$/M$_{star}$, left panel) and companion mass (M$_{Jup}$, right panel) vs separation (in au) of the four COPAINS sub-stellar candidates (star symbols) compared to those of previously known DI companions (filled circles). The colour bar shows the significativity of the TGAS-DR2 $\Delta\mu$ (SNR$_{\Delta \mu}$) used for the COPAINS selection. Objects with SNR$_{\Delta \mu}<3$ (or not in TGAS, and therefore without a measurement of the $\Delta\mu$) are shown in gray. }
    \label{fig:DIdmuDR2}
\end{figure*}

\section{Conclusions}\label{sec:conclusions}
We presented the results of the COPAINS survey, a search for companions to 25 stars selected using the COPAINS tool by \cite{Fontanive2019}, and the discovery of four new brown dwarf companions: HIP~21152~B, HIP~29724~B, HD~60584~B and HIP~63734~B. \\
The value of blind surveys of course lies in their ability to constrain the underlying planet and sub-stellar companion population, but comes with a high cost in terms of telescope time. 
On the other hand, surveys like COPAINS offer an undeniably efficient selection method, providing a much higher success rate with a considerably smaller time commitment. 
Moreover, the possibility to derive model independent mass estimates for the companions to accelerating stars also means that each new detection arising from surveys like COPAINS can be added to the currently scarce number of much needed benchmark objects, providing new crucial constraints the evolutionary models. Given the good agreement between the values of the masses of these companions obtained from the photometry with the model-independent ones based on the $\Delta\mu$, this work represents a considerable addition to the current benchmarks sample. \\
The combination with the FORECAST maps ensures a high detection rate even when the quality of the imaging data is not ideal, while at the same time further enhancing the sensitivity to lower mass companions.
Fig.~\ref{fig:DIdmuDR2} shows the mass ratio (left panel) and companion mass (right panel) vs separation of the COPAINS sub-stellar candidates compared to those of the known DI companions (data from \url{exoplanet.eu} updated on April 4$^{th}$, 2022). 
An estimate of the $\Delta\mu$ was possible for about 40\% of the known companions and about half of these (shown as coloured dots in Fig.~\ref{fig:DIdmuDR2}) would have been selected using COPAINS. This plot once again shows how a selection like the one provided by COPAINS can already lead to the detection of companions well in the planetary mass regime. \\
Catalogues like those by \cite{Brandt19} and \cite{Kervella2022} provide a more robust estimate of the accelerations, and therefore their use with COPAINS is likely to lead to an even more effective selection. 
Moreover, these acceleration catalogues can also be used to retrieve a time series of absolute astrometry which, combined with the imaging data, allows for a more detailed characterisation of the orbit, and thus of the dynamical mass \citep[see e.g.][]{Drimmel2021}.  The availability of Gaia-only accelerations in the upcoming full third {\it Gaia}Release (DR3) will allow for another great step further. It will in fact not only free the method from the boundaries so far imposed by the use of external catalogue as source of long term proper motion, but more importantly will allow for a significant improvement in terms of uncertainties, thus allowing future survey to truly focus on targets with accelerations caused by planetary mass companion.

\onecolumn 
\begin{table}
\renewcommand{\arraystretch}{1.4}
\caption{Sample characteristics. All the values of the parallax are from {\it Gaia}EDR3. The values of the age (Age$^{max}_{min}$) and stellar mass (Mass$^{max}_{min}$) are derived as described in Sec.~\ref{sec:ages} and \ref{sec:mstar} (see also Appendix~\ref{app:targets} for details on the age derivation for the single objects). The last column includes the observing period for which the target was originally proposed for and therefore the selection method applied, as detailed in Section~\ref{sec:selection}}
\resizebox{\linewidth}{!}{
\begin{tabular}{r|ll|ll|rrr|rr|l}
\hline
ID  & RA & Dec & Parallax & SpType  & $J$ & $H$ & $K$ & Age$^{max}_{min}$ & Mass$^{max}_{min}$ & Sel\\
    &   (hh:mm:ss)  & ($\degr$:$\arcmin$:$\arcsec$) & (mas) &  & mag & mag & mag & Myr &M$_{\odot}$ & \\
\hline\hline
HIP 15247 & 1:10:09.411 & -3:31:48.575 & 20.37 $\pm$ 0.03 & F6V & 6.457 & 6.209 & 6.099 & 45$^{150} _{30}$ & 1.23$^{+0.006} _{-0.003}$ & P104 \\
HIP 17439 & 3:44:09.0243 & -38:16:56.771 & 61.84 $\pm$ 0.02 & K2V & 5.462 & 5.088 & 4.934 & 700$^{900} _{500}$ & 0.87$^{+0.001} _{-0.001}$ & P104 \\
HIP 21152 & 4:32:04.7457 & 5:24:36.069 & 23.11 $\pm$ 0.03 & F5V & 5.593 & 5.385 & 5.333 & 625$^{700} _{600}$ & 1.44$^{+0.002} _{-0.002}$ & P104 \\
HIP 21317 & 4:34:35.2511 & 15:30:16.873 & 21.67 $\pm$ 0.02 & F8 & 6.745 & 6.552 & 6.445 & 625$^{700} _{600}$ & 1.12$^{+0.001} _{-0.001}$ & P100 \\
HIP 22506 & 4:50:35.4147 & -41:02:51.426 & 19.83 $\pm$ 0.05 & G9V & 7.438 & 7.099 & 6.876 & 50$^{125} _{40}$ & 0.96$^{+0.002} _{-0.009}$ & P104 \\
GJ 3346 & 5:19:59.4774 & -15:50:24.411 & 42.09 $\pm$ 0.02 & K3.5Vk: & 6.856 & 6.284 & 6.205 & 5400$^{6500} _{4300}$ & 0.74$^{+0.003} _{-0.003}$ & P100 \\
HIP 27441 & 5:48:36.7913 & -39:55:55.814 & 23.59 $\pm$ 0.01 & K0V & 7.407 & 7.061 & 6.912 & 250$^{350} _{150}$ & 0.90$^{+0.001} _{-0.001}$ & P100 \\
HIP 28474 & 6:00:41.2853 & -44:53:50.304 & 18.48 $\pm$ 0.01 & G8V & 7.73 & 7.433 & 7.321 & 42$^{125} _{35}$ & 0.96$^{+0.001} _{-0.020}$ & P104 \\
HIP 29724 & 6:15:38.8408 & -57:42:05.96 & 15.86 $\pm$ 0.01 & G2V & 7.715 & 7.447 & 7.346 & 150$^{250} _{100}$ & 1.04$^{+0.001} _{-0.001}$ & P104 \\
HIP 33690 & 6:59:59.8409 & -61:20:12.444 & 54.53 $\pm$ 0.01 & G9V & 5.459 & 5.097 & 4.987 & 650$^{800} _{500}$ & 0.94$^{+0.001} _{-0.002}$ & P104 \\
HD 57852 & 7:20:21.4547 & -52:18:42.709 & 28.63 $\pm$ 0.23 & F5V & 5.28 & 5.13 & 4.946 & 200$^{300} _{150}$ & 1.41$^{+0.006} _{-0.006}$ & P104 \\
HD 60584 & 7:34:18.6723 & -23:28:25.17 & 32.75 $\pm$ 0.04 & F5V+F6V & 5.03 & 4.9 & 4.773 & 1000$^{1700} _{300}$ & 1.35$^{+0.012} _{-0.016}$ & P100 \\
HIP 52462 & 10:43:28.4085 & -29:03:51.014 & 46.49 $\pm$ 0.02 & K1V & 6.176 & 5.77 & 5.66 & 170$^{250} _{120}$ & 0.86$^{+0.001} _{-0.001}$ & P104 \\
HIP 56153 & 11:30:35.5865 & -57:08:02.252 & 44.53 $\pm$ 0.02 & K3.5V(k) & 6.495 & 6.019 & 5.868 & 700$^{1000} _{500}$ & 0.79$^{+0.001} _{-0.001}$ & P102 \\
HIP 59726 & 12:14:57.6323 & -41:08:21.243 & 31.14 $\pm$ 0.02 & G5V & 6.31 & 6.008 & 5.925 & 700$^{1000} _{500}$ & 1.02$^{+0.002} _{-0.003}$ & P104 \\
HIP 61804 & 12:40:00.1103 & -17:41:03.596 & 16.20 $\pm$ 0.31 & G3V & 7.295 & 7.043 & 6.869 & 120$^{200} _{80}$ & 1.12$^{+0.009} _{-0.010}$ & P104 \\
HIP 63734 & 13:03:39.0733 & -16:20:11.414 & 18.49 $\pm$ 0.03 & F7/8V & 6.752 & 6.522 & 6.436 & 150$^{300} _{100}$ & 1.21$^{+0.006} _{-0.006}$ & P104 \\
HIP 63862 & 13:05:16.9255 & -50:51:23.776 & 21.06 $\pm$ 0.02 & G6V & 7.158 & 6.834 & 6.744 & 200$^{300} _{120}$ & 1.03$^{+0.001} _{-0.002}$ & P100 \\
HIP 71899 & 14:42:23.1023 & 21:17:35.386 & 21.94 $\pm$ 0.02 & F8 & 6.447 & 6.223 & 6.172 & 300$^{3000} _{100}$ & 1.18$^{+0.011} _{-0.012}$ & P102 \\
HIP 78549 & 16:02:13.5632 & -22:41:15.023 & 7.01 $\pm$ 0.03 & B9.5V & 7.037 & 7.038 & 6.97 & 11$^{12} _{4}$ & 2.61$^{+0.010} _{-0.014}$ & P104 \\
HIP 108912 & 22:03:42.303 & -60:26:14.87 & 23.38 $\pm$ 0.02 & G2V & 6.773 & 6.529 & 6.444 & 300$^{400} _{200}$ & 1.07$^{+0.002} _{-0.002}$ & P104 \\
HIP 112491 & 22:47:09.1673 & -32:40:30.874 & 36.43 $\pm$ 0.02 & G8V & 6.295 & 5.958 & 5.852 & 700$^{900} _{500}$ & 0.95$^{+0.001} _{-0.002}$ & P104 \\
HIP 112581 & 22:48:06.8092 & -37:45:23.989 & 26.14 $\pm$ 0.02 & G0VCH-0.3 & 6.415 & 6.172 & 6.1 & 600$^{800} _{450}$ & 1.09$^{+0.002} _{-0.003}$ & P104 \\
CD-69 2101 & 23:31:00.5329 & -69:05:09.756 & 33.03 $\pm$ 0.01 & K3V & 7.345 & 6.824 & 6.712 & 350$^{500} _{200}$ & 0.77$^{+0.001} _{-0.001}$ & P100 \\
HIP 116768 & 23:39:54.9912 & 9:40:38.345 & 14.05 $\pm$ 0.29 & A2m & 5.563 & 5.488 & 5.471 & 780$^{1040} _{520}$ & 1.88$^{+0.045} _{-0.062}$ & P104 \\
\hline\hline
\end{tabular}}\label{tab:master}
\end{table}

\begin{table}
\caption{Proper motion values from TGAS, DR2 and EDR3 and the resulting $\Delta\mu$ values obtained considering TGAS or Tycho-II as source for long term proper motions and DR2 or EDR3 as sources short-term proper motions.}
    
 \resizebox{\linewidth}{!}{
    \begin{tabular}{r|rrrr|rrrr|rrr}
 \hline\hline
 ID  & \multicolumn{4}{c|}{pmRA} & \multicolumn{4}{c|}{pmDE} & \multicolumn{3}{c}{$\Delta\mu$} \\
     & TYCHO-2 & TGAS & DR2 & EDR3        & TYCHO-2 &  TGAS & DR2 & EDR3       & TYC-TGAS & TGAS-DR2 & TGAS-DR3 \\
\hline
HIP 15247 & 77.70 $\pm$ 1.40 & 79.25 $\pm$ 0.04 & 82.10 $\pm$ 0.09 & 82.10 $\pm$ 0.03 & -45.80 $\pm$ 1.30 & -47.25 $\pm$ 0.03 & -49.08 $\pm$ 0.08 & -49.05 $\pm$ 0.03 & 2.13 $\pm$ 0.05 & 3.39 $\pm$ 0.09 & 3.36 $\pm$ 0.05 \\
HIP 17439 & 209.90 $\pm$ 1.20 & 209.07 $\pm$ 0.04 & 208.91 $\pm$ 0.03 & 209.06 $\pm$ 0.02 & 291.00 $\pm$ 1.20 & 289.26 $\pm$ 0.05 & 289.30 $\pm$ 0.05 & 289.32 $\pm$ 0.02 & 1.93 $\pm$ 0.05 & 0.16 $\pm$ 0.05 & 0.06 $\pm$ 0.05 \\
HIP 21152 & 113.20 $\pm$ 1.30 & 112.59 $\pm$ 0.04 & 112.50 $\pm$ 0.13 & 112.17 $\pm$ 0.03 & 5.40 $\pm$ 1.30 & 8.06 $\pm$ 0.02 & 7.58 $\pm$ 0.06 & 7.76 $\pm$ 0.02 & 2.73 $\pm$ 0.04 & 0.49 $\pm$ 0.06 & 0.51 $\pm$ 0.04 \\
HIP 21317 & 98.10 $\pm$ 1.00 & 101.12 $\pm$ 0.10 & 100.87 $\pm$ 0.11 & 100.96 $\pm$ 0.03 & -26.70 $\pm$ 1.10 & -26.74 $\pm$ 0.06 & -26.86 $\pm$ 0.07 & -26.81 $\pm$ 0.02 & 3.02 $\pm$ 0.10 & 0.28 $\pm$ 0.14 & 0.18 $\pm$ 0.10 \\
HIP 22506 & 37.10 $\pm$ 1.10 & 36.77 $\pm$ 0.04 & 36.45 $\pm$ 0.13 & 36.46 $\pm$ 0.06 & 69.90 $\pm$ 1.10 & 69.08 $\pm$ 0.05 & 68.07 $\pm$ 0.14 & 68.59 $\pm$ 0.07 & 0.88 $\pm$ 0.08 & 1.06 $\pm$ 0.15 & 0.58 $\pm$ 0.08 \\
GJ 3346 & 174.30 $\pm$ 1.30 & 174.02 $\pm$ 0.06 & 173.57 $\pm$ 0.05 & 173.70 $\pm$ 0.01 & 201.20 $\pm$ 1.40 & 206.39 $\pm$ 0.06 & 207.56 $\pm$ 0.06 & 207.71 $\pm$ 0.01 & 5.20 $\pm$ 0.07 & 1.25 $\pm$ 0.08 & 1.35 $\pm$ 0.07 \\
HIP 27441 & 26.50 $\pm$ 1.10 & 28.43 $\pm$ 0.05 & 28.46 $\pm$ 0.05 & 28.43 $\pm$ 0.01 & 71.10 $\pm$ 1.30 & 66.48 $\pm$ 0.05 & 66.21 $\pm$ 0.05 & 66.38 $\pm$ 0.01 & 5.01 $\pm$ 0.06 & 0.27 $\pm$ 0.07 & 0.11 $\pm$ 0.06 \\
HIP 28474 & 16.80 $\pm$ 1.00 & 17.52 $\pm$ 0.07 & 18.15 $\pm$ 0.05 & 18.22 $\pm$ 0.01 & 21.10 $\pm$ 1.10 & 23.09 $\pm$ 0.08 & 23.26 $\pm$ 0.05 & 23.47 $\pm$ 0.01 & 2.12 $\pm$ 0.07 & 0.65 $\pm$ 0.08 & 0.79 $\pm$ 0.07 \\
HIP 29724 & 30.20 $\pm$ 1.30 & 29.41 $\pm$ 0.07 & 27.67 $\pm$ 0.04 & 27.51 $\pm$ 0.02 & 48.30 $\pm$ 1.30 & 47.79 $\pm$ 0.07 & 48.46 $\pm$ 0.04 & 48.50 $\pm$ 0.02 & 0.94 $\pm$ 0.08 & 1.86 $\pm$ 0.08 & 2.02 $\pm$ 0.08 \\
HIP 33690 & -160.80 $\pm$ 1.30 & -161.89 $\pm$ 0.04 & -162.07 $\pm$ 0.05 & -161.87 $\pm$ 0.02 & 266.30 $\pm$ 1.60 & 264.87 $\pm$ 0.04 & 264.64 $\pm$ 0.04 & 264.84 $\pm$ 0.02 & 1.80 $\pm$ 0.04 & 0.29 $\pm$ 0.06 & 0.04 $\pm$ 0.04 \\
HD 57852 & -62.50 $\pm$ 1.20 & -36.88 $\pm$ 0.05 & -37.71 $\pm$ 0.59 & -37.05 $\pm$ 0.32 & 107.00 $\pm$ 1.50 & 146.69 $\pm$ 0.05 & 148.39 $\pm$ 0.56 & 146.29 $\pm$ 0.29 & 47.24 $\pm$ 0.30 & 1.89 $\pm$ 0.56 & 0.43 $\pm$ 0.30 \\
HD 60584 & -83.50 $\pm$ 1.10 & -88.05 $\pm$ 0.04 & -88.13 $\pm$ 0.07 & -88.13 $\pm$ 0.02 & -0.80 $\pm$ 1.00 & -0.59 $\pm$ 0.04 & -0.74 $\pm$ 0.09 & -0.86 $\pm$ 0.04 & 4.55 $\pm$ 0.05 & 0.17 $\pm$ 0.09 & 0.28 $\pm$ 0.05 \\
HIP 52462 & -212.50 $\pm$ 1.50 & -215.57 $\pm$ 0.03 & -215.54 $\pm$ 0.06 & -215.48 $\pm$ 0.01 & -50.00 $\pm$ 1.20 & -49.88 $\pm$ 0.04 & -50.12 $\pm$ 0.06 & -49.89 $\pm$ 0.02 & 3.08 $\pm$ 0.04 & 0.24 $\pm$ 0.07 & 0.09 $\pm$ 0.04 \\
HIP 56153 & -553.30 $\pm$ 2.00 & -555.35 $\pm$ 0.07 & -555.29 $\pm$ 0.17 & -555.33 $\pm$ 0.02 & 36.40 $\pm$ 1.50 & 36.13 $\pm$ 0.06 & 35.91 $\pm$ 0.22 & 36.13 $\pm$ 0.02 & 2.07 $\pm$ 0.07 & 0.23 $\pm$ 0.22 & 0.02 $\pm$ 0.07 \\
HIP 59726 & -312.80 $\pm$ 0.90 & -313.95 $\pm$ 0.04 & -313.90 $\pm$ 0.05 & -313.97 $\pm$ 0.02 & -78.50 $\pm$ 0.90 & -77.04 $\pm$ 0.03 & -77.17 $\pm$ 0.03 & -77.00 $\pm$ 0.01 & 1.86 $\pm$ 0.03 & 0.14 $\pm$ 0.05 & 0.04 $\pm$ 0.03 \\
HIP 61804 & -48.00 $\pm$ 1.30 & -45.70 $\pm$ 0.07 & -47.83 $\pm$ 0.38 & -46.64 $\pm$ 0.34 & -10.40 $\pm$ 1.40 & -11.60 $\pm$ 0.05 & -12.23 $\pm$ 0.32 & -11.15 $\pm$ 0.22 & 2.59 $\pm$ 0.33 & 2.22 $\pm$ 0.38 & 1.04 $\pm$ 0.33 \\
HIP 63734 & -108.70 $\pm$ 1.20 & -108.30 $\pm$ 0.05 & -108.48 $\pm$ 0.10 & -108.37 $\pm$ 0.03 & -28.90 $\pm$ 1.10 & -29.55 $\pm$ 0.03 & -29.31 $\pm$ 0.07 & -29.30 $\pm$ 0.02 & 0.76 $\pm$ 0.04 & 0.30 $\pm$ 0.09 & 0.26 $\pm$ 0.04 \\
HIP 63862 & -135.50 $\pm$ 1.10 & -134.34 $\pm$ 0.04 & -134.15 $\pm$ 0.11 & -134.34 $\pm$ 0.02 & -1.10 $\pm$ 1.10 & -4.45 $\pm$ 0.05 & -4.65 $\pm$ 0.08 & -4.45 $\pm$ 0.02 & 3.54 $\pm$ 0.05 & 0.28 $\pm$ 0.11 & 0.00 $\pm$ 0.05 \\
HIP 71899 & -108.00 $\pm$ 1.00 & -107.27 $\pm$ 0.03 & -108.39 $\pm$ 0.06 & -108.42 $\pm$ 0.01 & -31.50 $\pm$ 1.00 & -34.51 $\pm$ 0.04 & -35.21 $\pm$ 0.07 & -35.14 $\pm$ 0.02 & 3.10 $\pm$ 0.03 & 1.32 $\pm$ 0.07 & 1.31 $\pm$ 0.03 \\
HIP 78549 & -12.00 $\pm$ 1.40 & -12.54 $\pm$ 0.04 & -12.51 $\pm$ 0.11 & -12.53 $\pm$ 0.03 & -24.70 $\pm$ 1.40 & -23.13 $\pm$ 0.03 & -23.53 $\pm$ 0.06 & -23.65 $\pm$ 0.02 & 1.66 $\pm$ 0.03 & 0.40 $\pm$ 0.06 & 0.52 $\pm$ 0.03 \\
HIP 108912 & 105.30 $\pm$ 1.40 & 106.47 $\pm$ 0.05 & 106.27 $\pm$ 0.04 & 106.40 $\pm$ 0.01 & 2.10 $\pm$ 1.40 & 2.08 $\pm$ 0.05 & 2.20 $\pm$ 0.05 & 2.00 $\pm$ 0.02 & 1.17 $\pm$ 0.05 & 0.23 $\pm$ 0.07 & 0.11 $\pm$ 0.05 \\
HIP 112491 & 248.00 $\pm$ 1.30 & 247.31 $\pm$ 0.07 & 247.14 $\pm$ 0.08 & 247.28 $\pm$ 0.02 & -90.60 $\pm$ 1.20 & -91.24 $\pm$ 0.05 & -90.86 $\pm$ 0.06 & -91.25 $\pm$ 0.02 & 0.94 $\pm$ 0.07 & 0.42 $\pm$ 0.09 & 0.02 $\pm$ 0.07 \\
HIP 112581 & 158.30 $\pm$ 1.60 & 156.54 $\pm$ 0.05 & 155.64 $\pm$ 0.07 & 155.72 $\pm$ 0.02 & 1.90 $\pm$ 1.20 & 1.49 $\pm$ 0.04 & 1.71 $\pm$ 0.08 & 1.57 $\pm$ 0.02 & 1.81 $\pm$ 0.05 & 0.93 $\pm$ 0.08 & 0.82 $\pm$ 0.05 \\
CD-69 2101 & 186.90 $\pm$ 1.50 & 181.14 $\pm$ 0.63 & 185.55 $\pm$ 0.04 & 185.66 $\pm$ 0.01 & -124.50 $\pm$ 1.40 & -125.88 $\pm$ 0.75 & -126.36 $\pm$ 0.05 & -126.53 $\pm$ 0.02 & 5.92 $\pm$ 0.64 & 4.43 $\pm$ 0.64 & 4.56 $\pm$ 0.64 \\
HIP 116768 & 88.60 $\pm$ 1.50 & 88.17 $\pm$ 0.05 & 86.91 $\pm$ 0.36 & 87.50 $\pm$ 0.28 & -11.30 $\pm$ 1.50 & -10.20 $\pm$ 0.04 & -10.18 $\pm$ 0.28 & -10.50 $\pm$ 0.19 & 1.18 $\pm$ 0.28 & 1.26 $\pm$ 0.36 & 0.74 $\pm$ 0.28 \\
\hline\hline 
\end{tabular}}\label{tab:deltamu}
\end{table}

\begin{table}
\caption{Summary of VLT/SPHERE observations.}
\resizebox{\linewidth}{!}{
\begin{tabular}{lcccccccc}
\hline\hline
ID & OBS DATE & MJD & MODE & DITxNDIT & ND Filt & FoV rot & seeing & $\tau_0$ \\
\hline
HIP 15247   & 2019-10-03 & 58759.27718 & IRDIFS-EXT & 64x2 & ND 3.5 & 19.54 & 0.89 & 0.0021 \\
            & 2019-10-18 & 58774.26027 & IRDIFS-EXT & 64x2 & ND 3.5 & 2.13 & 1.03 & 0.0024 \\
            & 2019-10-26 & 58782.22189 & IRDIFS-EXT & 64x2 & ND 3.5 & 25.60 & 1.0 & 0.0026 \\
HIP 17439   & 2019-10-02 & 58758.30404 & IRDIFS-EXT & 32x6 & ND 3.5 & 49.90 & 1.06 & 0.0022 \\
HIP 21152   & 2019-10-26 & 58782.29206 & IRDIFS-EXT & 64x3 & ND 3.5 & 36.63 & 1.19 & 0.0032 \\
            & 2019-10-26 & 58782.29206 & IRDIFS-EXT & 64x3 & ND 3.5 & 36.63 & 1.19 & 0.0032 \\
HIP 21317   & 2019-03-03 & 58545.34589 & IRDIFS-EXT & 64x1 & ND 2.0 & 5.93 & 0.93 & 0.0043 \\
HIP 22506   & 2019-10-03 & 58759.31361 & IRDIFS-EXT & 64x3 & ND 3.5 & 31.72 & 0.98 & 0.0018 \\
GJ 3346     & 2018-01-29 & 58147.10689 & IRDIFS-EXT & 64x1 & ND 2.0 & 5.16 & 1.18 & 0.0039 \\
HIP 27441   & 2018-10-03 & 58394.32920 & IRDIFS-EXT & 64x1 & ND 2.0 & 6.63 & 1.22 & 0.003 \\
HIP 28474   & 2019-10-16 & 58772.26844 & IRDIFS-EXT & 64x3 & ND 3.5 & 31.85 & 1.2 & 0.002 \\
            & 2019-10-28 & 58784.31147 & IRDIFS-EXT & 64x3 & ND 3.5 & 37.70 & 1.01 & 0.0038 \\
HIP 29724   & 2019-10-12 & 58768.32106 & IRDIFS-EXT & 64x2 & ND 3.5 & 12.38 & 1.04 & 0.0028 \\
HIP 33690   & 2019-12-05 & 58822.24603 & IRDIFS-EXT & 64x3 & ND 3.5 & 19.75 & 0.93 & 0.0033 \\
HD 57852    & 2019-12-14 & 58831.24541 & IRDIFS-EXT & 64x3 & ND 3.5 & 27.11 & 0.93 & 0.0089 \\
HD 60584    & 2018-02-03 & 58152.24399 & IRDIFS-EXT & 64x1 & ND 3.5 & 0.84 & 1.01 & 0.0064 \\
            & 2018-10-08 & 58399.33920 & IRDIFS-EXT & 64x1 & ND 3.5 & 1.82 & 1.05 & 0.0024 \\
            & 2018-11-25 & 58447.22078 & IRDIFS-EXT & 64x1 & ND 3.5 & 0.82 & 0.91 & 0.0084 \\
            & 2019-12-16 & 58833.30235 & IRDIFS-EXT & 32x5 & ND 3.5 & 1.62 & 1.68 & 0.0027 \\
HIP 52462   & 2020-01-04 & 58852.23233 & IRDIFS-EXT & 32x6 & ND 3.5 & 6.28 & 1.22 & 0.0097 \\
HIP 56153   & 2019-02-03 & 58545.27470 & IRDIFS-EXT	& 16x4 & ND 2.0 & 0.41 & 1.05 & 0.005\\
HIP 59726   & 2020-01-08 & 58856.26851 & IRDIFS-EXT & 64x3 & ND 3.5 & 2.98 & 1.29 & 0.0015 \\
            & 2020-01-19 & 58867.24264 & IRDIFS-EXT & 64x3 & ND 3.5 & 11.37 & 1.02 & 0.0124 \\
HIP 61804   & 2020-02-22 & 58901.28462 & IRDIFS-EXT & 64x2 & ND 3.5 & 57.53 & 1.13 & 0.0051 \\
HIP 63734   & 2020-02-07 & 58886.27393 & IRDIFS-EXT & 64x3 & ND 3.5 & 6.36 & 1.15 & 0.0041 \\
HIP 63862   & 2018-02-04 & 58153.34083 & IRDIFS-EXT & 64x1 & ND 2.0 & 9.25 & 1.0 & 0.0079 \\
HIP 71899   & 2019-03-03 & 58545.34590 & IRDIFS-EXT	& 64x1 & ND 2.0	& 5.93 & 0.93 & 0.0043\\
HIP 78549   & 2020-03-24 & 58932.30214 & IRDIFS-EXT & 32x6 & ND 3.5 & 8.62 & 1.24 & 0.0042 \\
HIP 108912 & 2019-10-04 & 58760.06723 & IRDIFS-EXT & 64x3 & ND 3.5 & 20.45 & 1.25 & 0.0021 \\
HIP 112491  & 2019-10-02 & 58758.18845 & IRDIFS-EXT & 64x2 & ND 3.5 & 6.99 & 1.28 & 0.0009 \\
HIP 112581 & 2019-10-04 & 58760.13838 & IRDIFS-EXT & 64x3 & ND 1.0 & 27.74 & 0.81 & 0.003 \\
CD-69 2101 & 2017-10-01 & 58027.12834 & IRDIFS-EXT & 64x1 & ND 2.0 & 6.12 & 0.99 & 0.0034 \\
HIP 116768 & 2019-10-05 & 58761.11277 & IRDIFS-EXT & 64x2 & ND 3.5 & 16.74 & 0.8 & 0.0023 \\

\hline\hline
\end{tabular}}\label{tab:obslog}
\end{table}

\begin{table}
\caption{IFS contrast limits (expressed as $\Delta mag$) for all the available data sets.} 
\resizebox{\linewidth}{!}{
\begin{tabular}{ll|lllllllll}
\hline\hline
ID  & JD  & 52mas & 100mas & 200mas & 300mas & 400mas & 500mas & 600mas & 700mas & 800mas  \\
\hline
HIP 15247 & 03/10/2019 &  & 7.49 & 10.08 & 11.96 & 12.9 & 12.86 & 12.68 & 13.23 & 13.94 \\
HIP 15247 & 18/10/2019 &  & 6.04 & 9.52 & 9.82 & 10.44 & 11.66 & 11.61 & 10.59 & 12.25 \\
HIP 15247 & 26/10/2019 & 3.89 & 7.36 & 9.9 & 12.25 & 13.32 & 13.52 & 13.1 & 12.79 & 13.46 \\
HIP 17439 & 02/10/2019 & 5.17 & 8.03 & 11.76 & 13.92 & 14.09 & 13.95 & 14.83 & 15.26 & 14.35 \\
HIP 21152 & 26/10/2019 & 6.09 & 7.53 & 12.13 & 14.41 & 14.01 & 14.15 & 14.49 & 14.5 & 13.83 \\
HIP 21317 & 02/02/2018 &  & 6.85 & 10.24 & 11.03 & 11.9 & 12.25 & 11.41 & 12.29 & 11.88 \\
HIP 71899 & 03/03/2019 &  & 7.26 & 10.75 & 12.5 & 12.68 & 13.49 & 12.81 & 13.26 & 13.24 \\
HIP 22506 & 03/10/2019 &  & 7.88 & 11.01 & 13.88 & 13.47 & 13.59 & 13.25 & 12.66 & 10.1 \\
GJ 3346   & 29/01/2018 &  & 7.17 & 11.55 & 12.22 & 12.11 & 12.86 & 13.03 & 12.86 & 12.72 \\
HIP 27441 & 03/10/2018 &  & 5.54 & 9.55 & 10.7 & 11.11 & 11.07 & 11.54 & 12.65 & 12.14 \\
HIP 28474 & 16/10/2019 & 3.36 & 7.76 & 10.16 & 11.98 & 11.73 & 11.69 & 12.33 & 10.31 & 11.44 \\
HIP 28474 & 28/10/2019 & 3.77 & 6.47 & 10.0 & 12.05 & 12.09 & 10.96 & 13.27 & 9.44 & 12.33 \\
HIP 29724 & 12/10/2019 & 3.91 & 8.66 & 9.88 & 10.86 & 11.41 & 11.71 & 11.97 & 12.0 & 11.8 \\
HIP 33690 & 05/12/2019 & 6.42 & 7.07 & 11.72 & 13.96 & 14.54 & 14.58 & 14.58 & 15.04 & 14.75 \\
HD 57852 & 14/12/2019 & 4.62 & 7.37 & 11.8 & 13.92 & 14.5 & 14.16 & 14.74 & 14.73 & 14.63 \\
HD 60584 & 03/02/2018 &  & 6.34 & 10.35 & 10.87 & 12.68 & 11.47 & 13.02 & 11.06 & 11.82 \\
HD 60584 & 08/10/2018 &  & 7.0 & 11.32 & 11.74 & 12.3 & 12.38 & 12.15 & 12.32 & 12.2 \\
HD 60584 & 25/11/2018 &  & 7.24 & 10.77 & 11.41 & 11.94 & 12.62 & 13.56 & 11.99 & 12.31 \\
HD 60584 & 16/12/2019 &  & 7.56 & 10.12 & 11.18 & 11.97 & 12.83 & 12.95 & 12.87 & 13.0 \\
HIP 52462 & 04/01/2020 &  & 7.47 & 10.86 & 11.81 & 13.16 & 13.33 & 12.69 & 13.26 & 14.0 \\
HIP 56153 & 02/03/2019 &  & 6.98 & 10.47 & 11.44 & 12.08 & 12.31 & 12.29 & 11.75 & 10.54 \\
HIP 59726 & 08/01/2020 &  & 3.56 & 7.13 & 8.5 & 10.84 & 12.17 & 10.41 & 11.1 & 10.27 \\
HIP 61804 & 22/02/2020 & 3.74 & 7.68 & 11.5 & 12.84 & 13.16 & 13.69 & 13.79 & 13.17 & 13.22 \\
HIP 63734 & 07/02/2020 &  & 7.12 & 10.88 & 11.91 & 12.44 & 13.36 & 13.33 & 12.99 & 11.74 \\
HIP 63862 & 04/02/2018 &  & 4.28 & 10.21 & 12.05 & 12.39 & 12.85 & 12.91 & 13.29 & 13.06 \\
HIP 78549 & 24/03/2020 &  & 5.9 & 8.76 & 10.27 & 11.7 & 11.59 & 12.03 & 11.57 & 11.15 \\
HIP 108912 & 04/10/2019 & 4.33 & 8.07 & 10.08 & 12.59 & 12.95 & 12.5 & 13.51 & 13.37 & 13.34 \\
HIP 112491 & 02/10/2019 &  &  &  &  &  &  &  &  &  \\
HIP 112581 & 04/10/2019 & 5.69 & 7.73 & 10.45 & 11.77 & 12.52 & 13.0 & 11.98 & 11.6 & 10.53 \\
CD-69 2101 & 01/10/2017 &  & 5.65 & 9.37 & 11.18 & 11.08 & 11.69 & 12.26 & 12.32 & 12.36 \\
HIP 116768 & 05/10/2019 & 5.33 & 7.04 & 10.92 & 12.98 & 13.23 & 14.06 & 13.49 & 14.62 & 14.52 \\
 \hline \hline
\end{tabular}}\label{tab:ifs_ccurves}
\end{table}

\begin{table}
\caption{SPHERE astrometry and photometry for all the candidate companions detected in our sample. Each epoch is reported separately. The last two column show the status of the candidate and the probability of finding a background contami   t at the given separation and contrast as a function of galactic coordinates, derived as described in Sec.~\ref{sec:common_pm}.  }
\resizebox{\linewidth}{!}{
\begin{tabular}{r|l|ll|ll|ll|ll}

\hline \hline
ID  & Obs. Date & rho & PA & \multicolumn{2}{c|}{IFS Photometry} & \multicolumn{2}{c|}{IRDIS Photometry}&   Status & Bkg Prob\\
    & (MJD) & (mas) & (\degr) &  \multicolumn{1}{|c}{$\Delta Y$} & \multicolumn{1}{c|}{$\Delta J$}  & \multicolumn{1}{c}{$\Delta K_1$} & $\Delta K_2$ & & \\
\hline \hline
HIP 15247   & 58759.2772 & 810.10 $\pm$ 1.50 & 121.30 $\pm$ 0.11 &     &     & 2.02 $\pm$ 0.03 & 2.15 $\pm$ 0.12 & C1 & 1.7910$^{-7}$ \\
            & 58774.2603 & 810.10 $\pm$ 1.50 & 121.30 $\pm$ 0.11 &     & 3.56 $\pm$ 0.50 & 2.15 $\pm$ 0.03 & 2.16 $\pm$ 0.14 & &  \\
            & 58782.2219 & 808.80 $\pm$ 1.50 & 121.50 $\pm$ 0.10 &     & 3.65 $\pm$ 0.50 & 2.11 $\pm$ 0.03 & 2.34 $\pm$ 0.22 &  &  \\ 
GJ 3346     & 58147.1069 & 3665.02 $\pm$ 2.27 & 348.31 $\pm$ 1.42 &     &     & 7.84 $\pm$ 1.63 & 7.79 $\pm$ 0.06 & C1 & 2.05 10$^{-4}$ \\
HIP 21152   & 58782.2921 & 422.40 $\pm$ 1.50 & 217.06 $\pm$ 0.20 & 11.65 $\pm$ 0.50 & 11.92 $\pm$ 0.50 & 10.82 $\pm$ 0.04 & 10.86 $\pm$ 0.05 & C0 & 8.17 10$^{-7}$ \\
HIP 28474   & 58772.2684 & 696.80 $\pm$ 1.50 & 75.42 $\pm$ 0.12 & 3.46 $\pm$ 0.50 & 3.75 $\pm$ 0.50 & 3.71 $\pm$ 0.03 & 3.63 $\pm$ 0.02 & C1 & 4.05 10$^{-6}$ \\
            & 58784.3115 & 694.40 $\pm$ 1.50 & 75.44 $\pm$ 0.12 & 3.77 $\pm$ 0.50 & 3.61 $\pm$ 0.50 &     &     & &  \\
HIP 29724   & 58768.3211 & 99.90 $\pm$ 1.50 & 214.00 $\pm$ 0.86 & 5.81 $\pm$ 0.50 & 6.36 $\pm$ 0.50 & 5.63 $\pm$ 0.20 & 6.00 $\pm$ 0.70 & C0 & 6.57 10$^{-7}$ \\
HIP 33690   & 58822.246 & 3117.57 $\pm$ 1.10 & 239.87 $\pm$ 0.87 &     &     & 5.41 $\pm$ 0.01 & 7.70 $\pm$ 0.17 & B2 &  \\
HD 57852    & 58831.2454 & 4792.45 $\pm$ 6.50 & 20.18 $\pm$ 0.30 &     &     & 20.18 $\pm$ 3.67 & 15.88 $\pm$ 1.81 & B2 &  \\
HD 60584 CC1 & 58447.2208 & 543.00 $\pm$ 5.00 & 52.50 $\pm$ 0.50 &     & 12.70 $\pm$ 0.50 &     &     & L & 8.34 10$^{-4}$ \\
HD 60584 CC2 & 58152.244 & 4475.87 $\pm$ 5.24 & 332.00 $\pm$ 0.06 &     &     & 10.05 $\pm$ 0.01 & 10.00 $\pm$ 0.01 & B1 &  \\
            & 58399.3392 & 4459.78 $\pm$ 2.21 & 332.20 $\pm$ 0.03 &     &     & 10.41 $\pm$ 0.09 & 10.35 $\pm$ 0.06 & &  \\
            & 58447.2208 & 4461.29 $\pm$ 1.23 & 332.51 $\pm$ 0.02 &     &     & 9.94 $\pm$ 0.04 & 10.65 $\pm$ 0.02 &  &  \\
HIP 63734   & 58886.2739 & 555.00 $\pm$ 2.00 & 329.60 $\pm$ 0.20 & 11.26 $\pm$ 0.50 & 12.08 $\pm$ 0.50 &     &     & L &  5.30 10$^{-5}$\\
HIP 63862   & 58153.3408 & 4815.57 $\pm$ 5.67 & 38.62 $\pm$ 0.06 &     &     & 12.72 $\pm$ 0.11 & 12.33 $\pm$ 0.45 & B2 &  \\
HIP 71899   & 58545.3459 & 1115.91 $\pm$ 4.50 & 241.69 $\pm$ 0.05 &     &     & 3.72 $\pm$ 0.01 & 3.98 $\pm$ 0.11 & C1 & 3.39 10$^{-6}$ \\
HIP 78549   & 58932.3021 & 333.20 $\pm$ 1.50 & 183.40 $\pm$ 0.26 & 4.24 $\pm$ 0.50 & 4.68 $\pm$ 0.50 & 9.58 $\pm$ 0.65 & 12.37 $\pm$ 1.35 & C1 & 2.52 10$^{-6}$ \\
HIP 112581 & 58760.1384 & 736.80 $\pm$ 1.50 & 286.30 $\pm$ 0.12 & 4.71 $\pm$ 0.50 & 5.00 $\pm$ 0.50 & 5.48 $\pm$ 0.01 & 5.32 $\pm$ 0.01 & C0 & 2.17 10$^{-6}$ \\
\hline\hline
\end{tabular}}
\\
\textbf{Status:} \textit{C0} = bound companion based on statistical arguments ; \textit{C1} = bound companion based on additional epochs from other works; \textit{B0}=	background based on statistical arguments; \textit{B1} =	background based on follow-up from this work; \textit{B2} =	background based on based on additional epochs from other works; \textit{U} = L = Low SNR detection (see Sec.~\ref{sec:forecast})\label{tab:photoastro}
\end{table}

\begin{table}
   \caption{Gaia astrometry and photometry of companions retrieved in EGDR3, including additional companions outside SPHERE FoV. Separations and position angles were derived using the positions from Gaia EDR3, when available. Further details about the known systems can be found in Appendix~\ref{app:targets}. The ID of the target is reported in the first column to be consistent with the rest of the tables in the paper. The Gaia ID, and the values of the parallax and proper motion reported are those retrieved in EDR3 for the companions. Although the companions of HIP 15247 and HD 129501 were detected by Gaia, no astrometric solution was available in EDR3 (hence the blank fields). 
   } 
\resizebox{\linewidth}{!}{   \begin{tabular}{rl|lll|lll} 
   \hline \hline
      ID$_A$ 	& Gaia EDR3 ID$_B$ &   parallax & \multicolumn{2}{c|}{Proper Motion} &$\Delta$mag  & separation    & PA 	\\
                &       		 &  (mas)	    & RA (mas/yr) 	& DEC (mas/yr)	    &Gaia G band  &  (arcsec)		& (\degr) \\
      \hline\hline  
HIP 15247 & Gaia EDR3 3261733202649353216 &  &  &  &  & 0.84 & 120.56 \\
HIP 24874 & Gaia EDR3 2983256662868370048 & 42.240 $\pm$ 0.035 & 182.685 $\pm$ 0.031 & 216.203 $\pm$ 0.032 & 14.331 & 3.65 & 347.98 \\
HIP 27441 & Gaia EDR3 4805207967655791488 & 23.675 $\pm$ 0.013 &  31.712 $\pm$ 0.014 &  65.198 $\pm$ 0.014 & 12.303 & 33.98 & 113.54 \\
 HD 60584 & Gaia EDR3 5618420137803146240 & 32.792 $\pm$ 0.035 & -87.258 $\pm$ 0.016 & -11.556 $\pm$ 0.036 & 5.758 & 9.95 & 117.61 \\
 HD 57852 & Gaia EDR3 5492026740698525696 & 29.476 $\pm$ 0.449 & -31.762 $\pm$ 0.546 & 138.163 $\pm$ 0.567 & 6.465 & 9.05 & 26.72 \\
HIP 71899 & Gaia EDR3 1241384331822285184 &  &  &  & 12.936 & 1.11 & 240.59 \\
\hline 
\hline 
\end{tabular}}\label{tab:Gaia_bin}
\end{table}

\begin{table}
\caption{Complete list of all the astrometric data for the detected candidate companions for which more than one epoch was available, with the appropriate references listed in the last column.  }
\resizebox{0.8\linewidth}{!}{
\begin{tabular}{lr|rr|lr|l}
\hline\hline
ID & Obs.Date       & rho   & $e_{rho}$ & PA        & $e_{PA}$  & Ref.\\
   & (MJD-245000)   & (mas) & (mas)     & (\degr) & (\degr) &    \\
\hline\hline
HIP 15247   	& 55576.4 & 0.879 & 0.003 & 121.15 & 2.80 & Hartkopf+2012 \\
            		& 55894.0 & 0.900 & 0.010 & 122.80 & 0.10 & Galicher+2016 \\
            		& 55916.0 & 0.870 & 0.010 & 118.67 & 0.49 & Meshkat+2015 \\
            		& 56263.9 & 0.872 & 0.006 & 121.20 & 5.70 & Tokovinin+2014 \\
            		& 56936.7 & 0.844 & 0.003 & 119.80 & 2.10 & Tokovinin+2015 \\
            		& 57388.0 & 0.837 & 0.001 & 120.55 & 0.02 & EDR3 \\
            		& 57738.3 & 0.830 & 0.001 & 119.80 & 0.90 & Tokovinin+ 2018 \\
            		& 58759.0 & 0.796 & 0.004 & 120.71 & 0.05 & This Work \\
            		& 58774.0 & 0.792 & 0.004 & 120.89 & 0.05 & This Work \\
            		& 58782.0 & 0.793 & 0.004 & 120.77 & 0.05 & This Work \\
GJ 3346     	& 57205.5 & 3.647 & 0.001 & 347.89 & 0.02 & DR2 \\
            		& 57388.0 & 3.651 & 0.001 & 347.98 & 0.02 & EDR3 \\
            		& 58147.0 & 3.665 & 0.002 & 348.30 & 0.07 & This Work \\
HIP 28474   	& 53122.0 & 0.613 & 0.003 &   61.70 & 0.20 & Chauvin+2015 \\
            		& 58501.9 & 0.693 & 0.003 &   74.20 & 2.90 & Tokovinin+2020 \\
            		& 58771.0 & 0.697 & 0.005 &   74.73 & 0.24 & This Work \\
HIP 33690   	& 54842.9 & 4.479 & 0.009 & 286.20 & 0.20 & Wahhaj+2013 \\
            		& 55200.6 & 4.302 & 0.009 & 282.90 & 0.20 & Wahhaj+2013 \\
            		& 58832.0 & 3.117 & 0.007 & 239.85 & 0.01 & This Work \\
HD 57852    	& 55157.0 & 3.301 & 0.020 & 148.90 & 0.30 & Chauvin+2015 \\
            		& 55243.0 & 3.365 & 0.013 & 148.70 & 0.20 & Chauvin+2015 \\
            		& 55592.0 & 3.465 & 0.009 & 149.80 & 0.20 & Chauvin+2015 \\
            		& 58831.0 & 4.962 & 0.006 & 154.73 & 0.12 & This Work \\
HD 60584    	& 58152.0 & 4.476 & 0.005 & 331.10 & 0.06 & This Work \\
            		& 58399.0 & 4.460 & 0.002 & 332.20 & 0.03 & This Work \\
            		& 58447.0 & 4.461 & 0.001 & 332.51 & 0.02 & This Work \\
HIP 63862   	& 55246.0 & 4.231 & 0.016 &   28.00 & 0.20 & Chauvin+2015 \\
            		& 55744.0 & 4.315 & 0.005 &   30.70 & 0.10 & Chauvin+2015 \\
           		& 58153.0 & 4.815 & 0.006 &   38.62 & 0.05 & This Work \\
HIP 71899   	& 57205.5 & 1.094 & 0.001 & 236.78 & 0.02 & DR2 \\
            		& 57388.0 & 1.108 & 0.001 & 240.59 & 0.02 & EDR3 \\
            		& 58545.0 & 1.116 & 0.004 & 241.69 & 0.05 & This Work \\
HIP 78549   	& 54645.0 & 0.325 & 0.001 & 189.06 & 0.19 & Lafreniere+2014 \\
            		& 58932.0 & 0.326 & 0.007 & 183.17 & 1.08 & This Work \\
\hline\hline 
\end{tabular}}\label{tab:multi_epoch}
\end{table}

\begin{table}
\caption{Summary of the characteristics of the comoving companions. If  more than one SHINE epoch was available, only the separation ($\rho$) and position angle (PA) from the first epoch are reported, information on the single measurements for these objects can be found in Tab.~\ref{tab:photoastro}.\\ 
Except for GJ~3346, for which we used the values from \citet{bonavita2020b}, the masses of the primaries ($M_A$) were derived as described in Sec.~\ref{sec:mstar}. 
The same is true for $M_B$(phot), which is the value of the secondary mass obtained from the photometry, using the COND models \citep{baraffe2003}. $M_B$($\Delta\mu$) is instead the value of the secondary mass inferred using FORECAST (see Sec.~\ref{sec:forecast} for details).}
 \resizebox{0.9\linewidth}{!}{   \begin{tabular}{l|rr|r|lll|l}
\hline
\hline 
\multicolumn{1}{c|}{ID}  & \multicolumn{2}{c|}{Separation}    & \multicolumn{1}{c|}{PA}    &  \multicolumn{1}{c}{$M_A$}           & \multicolumn{1}{c}{$M_B$(phot)}   & \multicolumn{1}{c|}{$M_B$($\Delta\mu$)} & Notes \\
    & \multicolumn{1}{c}{(mas)} & \multicolumn{1}{c|}{(au)}    & \multicolumn{1}{c|}{(deg)}  &  \multicolumn{1}{c}{($M_{\odot}$)}    & \multicolumn{1}{c}{($M_{\odot}$)} & \multicolumn{1}{c|}{($M_{\odot}$)}     & \\
\hline
HIP 15247 & 810.10 $\pm$ 1.50 & 39.78 $\pm$ 0.09 & 121.30 $\pm$ 0.11 & 1.228$\pm$0.004 & 0.680$\pm$0.015 & 0.668$\pm$0.369 &  \\
HIP 21152 & 422.40 $\pm$ 1.50 & 18.28 $\pm$ 0.07 & 217.06 $\pm$ 0.20 & 1.442$\pm$0.002 & 0.032$\pm$0.005 & 0.021$\pm$0.007 &  \\
GJ 3346 & 3665.02 $\pm$ 2.27 & 87.07 $\pm$ 0.06 & 348.31 $\pm$ 1.42 & 0.683$\pm$0.018 & 0.580$\pm$0.010 & 0.611$\pm$0.339 & WD \\
HIP 28474 & 696.80 $\pm$ 1.50 & 37.71 $\pm$ 0.08 & 75.42 $\pm$ 0.12 & 0.958$\pm$0.011 & 0.139$\pm$0.038 & 0.152$\pm$0.085 &  \\
HIP 29724 & 99.90 $\pm$ 1.50 & 6.30 $\pm$ 0.09 & 214.00 $\pm$ 0.86 & 1.044$\pm$0.001 & 0.063$\pm$0.008 & 0.067$\pm$0.021 &  \\
HD 60584 & 543.00 $\pm$ 5.00 & 16.58 $\pm$ 0.15 & 232.50 $\pm$ 0.50 & 1.352$\pm$0.014 & 0.028$\pm$0.009 & 0.008$\pm$0.003 & Low SNR \\
HIP 63734 & 555.00 $\pm$ 2.00 & 30.02 $\pm$ 0.12 & 329.60 $\pm$ 0.20 & 1.211$\pm$0.006 & 0.011$\pm$0.003 & 0.032$\pm$0.020 &  Low SNR \\
HIP 71899 & 1115.91 $\pm$ 4.50 & 50.85 $\pm$ 0.21 & 241.69 $\pm$ 0.05 & 1.178$\pm$0.012 & 0.390$\pm$0.005 & 0.393$\pm$0.218 &  \\
HIP 78549 & 333.20 $\pm$ 1.50 & 47.52 $\pm$ 0.30 & 183.40 $\pm$ 0.26 & 2.614$\pm$0.012 & 0.284$\pm$0.064 & 0.420$\pm$0.241 &  \\
HIP 112581 & 736.80 $\pm$ 1.50 & 28.19 $\pm$ 0.06 & 286.30 $\pm$ 0.12 & 1.091$\pm$0.003 & 0.144$\pm$0.050 & 0.062$\pm$0.035 &  \\
\hline\hline
\end{tabular}}\label{tab:sys_char}
\end{table}

\twocolumn

\section*{Data Availability}
The data used for this work are available through the ESO Science Archive Facility (\url{http://archive.eso.org/cms.html}). 

\bibliographystyle{mnras}
\bibliography{biblio}

\appendix
\section{Notes on individual targets}
\label{app:targets}

\medskip
\noindent{\bf  HIP 15247}
F6V star, classified as member of Tuc-Hor in several works \citep[e.g., ][]{zuckerman2004}
It was first spatially resolved into a close pair by \citet{hartkopf2012}.
Additional astrometric observations by \citet{tokovinin2014,tokovinin2016,tokovinin2018,riddle2015,galicher2016,meshkat2015} allows to constrain the binary orbit. The secondary is also detected in {\it Gaia}EDR3.
Significant RV variability (RV rms 6.6 km/s, 3 epochs over 850 days)
is reported by \citet{nordstrom2004}. Single-epoch (Feb 2003) RV by \citet{white2007} (9.1$\pm$0.8  km/s) is close to the \citet{nordstrom2004} mean value (7.2 km/s).
The three determinations by \citet{zuniga2021} (mean 17.78$\pm$0.38 km/s, span 16 days in July-August 2012) are instead off by more than 10 km/s with respect to \citet{nordstrom2004} mean value. These variations are larger than the expected ones caused by the known companion, making likely that the system is actually triple.
When using BANYAN $\Sigma$, the probability of Tuc-Hor membership largely depends on the adopted RV (high when adopting
\citet{nordstrom2004} and null for \citet{zuniga2021} RV). 
Li EW \citep{white2007} is more similar to that of Pleiades members, but compatible with Tuc-Hor age considering the dispersion of individual members, while the  X-ray luminosity is close to the median value of Tuc-Hor members of similar colors.
Considering the uncertainty in the group assignment due to multiplicity, we adopt Tuc-Hor membership, but allowing upper limit to stellar age up to the Pleiades age.
For this age, the mass of the secondary  results of 0.68 Msun.

\medskip
\noindent{\bf  HIP 17439 = HD 23484}
Field object with age indicators compatible with an age close to the Hyades. \citet{Raghavan2010} report tentative RV variability but the star results to have roughly constant RV from \citet{nordstrom2004} data (rms 0.1 km/s, N=4 over 2.8 years), \citet{soubiran2018} (rms 20 m/s from 5 CORALIE spectra over 13 years), and \citet{wittenmyer2015} (rms 14 m/s from 19 UCLES/AAT spectra over 8 years),
with the mean values of the first two of these works in agreement to better than 1 km/s.
The star has an extended debris disk, spatially resolved by Herschel
\citep{ertel2014}. Their modelling supports the presence of a two-component disk (belts at 29 and 90 au), possibly with a gap cleared by a massive planet, although a unique very wide belt can not be ruled out. At the expected location of such hypothetical planet ($\sim$ 60-80 au), the $\Delta \mu$ signature (albeit marginally significant) predict a mass of about 6-8 Mjup, which would place it below the observed detection limits for our images. RV data do not provide significant constraints at such wide separations.
The disk is not revealed in scattered light in our SPHERE images. 

\medskip
\noindent{\bf  HD 28736 = HIP 21152}
Member of Hyades from several works in the literature and confirmed with BANYAN $\Sigma$ analysis with updated kinematic parameters.
A new brown dwarf companion is detected in this study, as discussed in Sec.~\ref{sec:hd28736}.

\medskip
\noindent{\bf  HD 28992 = HIP 21152 = TYC 1266-278-1 }
Member of Hyades from several works in the literature and confirmed with BANYAN $\Sigma$ analysis with updated kinematic parameters.
RV monitoring from \citet{paulson2004} (rms 21 m/s, 11 measurements, baseline 5.2 yr with HIRES/Keck) and \citet{soubiran2018}
(rms 39 m/s, 11 measurements, baseline 2.8 yr with SOPHIE/OHP) rule out massive companions at moderately close separation.
The absolute RV by \citet{nordstrom2004} agrees within error with the recent ones by \citet{soubiran2018} and \citet{GaiaDR2}.

\medskip
\noindent{\bf  HIP 22506 = HD 31026 = TYC 7589-1186-1 }
Li EW \citep{sacy}, rotation period \citep{kiraga2012}, and X-ray emission are marginally compatible with the distributions of Pleiades (125 Myr) and Argus (50 Myr) members, but intermediate between the mean loci of these groups.
There are clear indications of RV variability (peak-to-valley 17 km/s) from sparse RV determinations available in the literature \citep{nordstrom2004,sacy,GaiaDR2}.
The assignment to known groups depends on the adopted RV. We obtain
high membership probability for Argus when adopting {\it Gaia}RV or without RV, and poor match for \citet{nordstrom2004,sacy} RV.
The expected RV to optimize the Argus membership is very close to the mean values of the three determinations.
We then adopt the Argus with upper limit at Pleiades age $50^{+75}_{-10}$ Myr.
Our SPHERE images do not reveal any candidate. It is possible 
that the $\Delta \mu$ signature is due to the unseen spectroscopic companion.

\medskip
\noindent{\bf  GJ 3346 = HIP 24874  = HD 34865 = TYC 5902-586-1 }\\
Star with white dwarf companion, discovered and characterized as part of the present project. Previously published in \citet{bonavita2020b}. It was included in the sample as it appears as a young star with moderately high activity level, due to accretion of material lost by the WD progenitor.

\medskip
\noindent{\bf HIP 27441 = HD 39126 = TYC 7601-371-1}
Field object. Age indicators consistently provide an age of 250$\pm$100 Myr.
The star has a wide common proper motion companion, 2MASS J05483951-3956087, at 34" (1400 au at the distance of the star).
From colors and absolute magnitudes, the companion is expected to be a M2 star.

\medskip
\noindent{\bf HIP 28474 =  HD 41071 = RT Pic}
The star is classified as member of Columba association in several works and membership is confirmed by our analysis with updated {\it Gaia}inputs. Age indicators support a young age although Li EW and rotation period are in mild disagreement with the typical values of Tuc-Hor and Columba members and more similar to Pleiades and AB Dor ones.
For this reason, we adopt Columba age but with upper limit at the age of the Pleiades. 
The classification as eclipsing binary with 2.6 mag deep eclipses \citep{malkov2006} appears spurious, as there are no indications of them in the TESS light curve (4 sectors) and the RV results constant over decades at km/s level \citep{nordstrom2004,GaiaDR2} and at few tens of m/s level over three months from HARPS data.
A close stellar companion was identified by \citet{Chauvin2010} and further observed by \citet{tokovinin2020}.  
It is also detected in our data (mass from photometry 0.16 Msun) and is the responsible of the $\Delta \mu$ signature.

\medskip
\noindent{\bf  HIP 29724 = HD 43976}
Young field object, not associated to any known moving group.
Li EW and X ray emission are close to the mean locus of the Pleiades
while the rotation period from TESS suggests a slightly older age.
A new substellar companion is detected in this study  (see Sec.~\ref{sec:hip29724} for details).

\medskip
\noindent{\bf  HIP 33690 = HD 53143}
Star with resolved debris disk \citep{kalas2008}.
The star is flagged as a possible member of IC2391/Argus in
some literature works \citep[e.g., ][]{Nielsen19}.
However, BANYAN analysis with updated {\it Gaia}parameters return 0\% membership probability for this group and no significant probability for other known MGs.
Furthermore, the non detection of Lithium \citep{torres2000} clearly rules out the young age of IC2391/Argus. The Li non-detection and the other age indicators are fully compatible with an age similar to the Hyades.
The companion candidate seen in our images and previously identified by \citet{wahhaj2013} is a background object.

\medskip
\noindent{\bf  HD 57852 = HIP 35564 = HR 2813 = TYC 8132-2112-1}
Bona-fide member of Carina-Near MG, confirmed with BANYAN $\Sigma$ analysis using the updated kinematic parameters. The star has highly significant RV variability \citep{andersen1983,pribulla2014,Desidera15,GaiaDR2},
although without orbital solution. Peak-to-valley RV range for the available data is 13 km/s.
The longest-period plausible orbital solution consistent with the data yields a period of 994 d with moderately high eccentricity. Shorter periods are also possible considering the poor time sampling.
The spectroscopic component is likely the responsible of the
$\Delta \mu$ signature and of the large RUWE in {\it Gaia}EDR3 (8.69), and it is expected to be at a separation too close to the star to be detected in SPHERE images.
HD 57852 has a wide companion (HD 57853) at 9", which is itself a spectroscopic binary with three components with total mass 2.4 Msun
\citep{saar1990,desidera2006a}
The system is then a hierarchical quintuple.
The source detected at 4.9" and previously identified by \citet{Chauvin2015} is
a background object.

\medskip
\noindent{\bf n Pup A = HD 60584 = TYC 6539-3802-1}
Mid-F type star with a nearly twin wide companion (n Pup B = HD 60585 at 9.9"). 
\citet{katoh2018} found the primary to an SB1 with period 366 days, eccentricity 0.49, RV semi-amplitude 12 km/s,  corresponding to a minimum mass of 0.5 Msun.
This orbital solution and variability above 1.6 km/s (peak-to-valley) are not supported by HARPS RV \citep{lagrange2009,Trifonov2020}.
The HARPS RV variability appears to be dominated by short-term variations (intra-night and from contiguous nights) with limited variability on longer timescales.
The star is an X-ray source, but considering the very limited age dependence of X-ray emission for mid-F stars we rely only on isochrone fitting for our age estimate.
The identification of a faint BD candidate, compatible with the $\Delta \mu$ signature, is discussed in Sect. \ref{sec:tyc6539-HIP6734}. The source at 4.4" is a background object.

\medskip
\noindent{\bf HIP 48341 = HD 85364 = HR 3899 = 6 Sex = TYC 4899-1904-1}
Early type star, identified as a possible member of UMa group by \citet{soderblom1993,king2003}. \citet{chupina2006} classified it as a member of corona surrounding the UMa nucleus.
Kinematic analysis using BANYAN returns a field object classification.
Independently on the kinematic assignment, isochrone fitting yields an age of 660$\pm$260, formally compatible with Ursa Major.
The star has also IR excess indicating the presence of a debris disk \citep{chen2014}.

\medskip
\noindent{\bf  HIP 52462 = HD 92945 = V419 Hya}
Star with spatially resolved debris disk \citep{golimowski2011,marino2019}.
The stellar properties and the presence of companions combining imaging, RV, and astrometric signatures were recently investigated by \citet{mesa2021}.

\medskip
\noindent{\bf  HIP 59726 = HD 106489}
Field object with age similar to the Hyades.
The RV results constant within 0.5 km/s from few literature measurements spanning several decades.

\medskip
\noindent{\bf  HIP 61804}
Young star, not associated to any known moving group.
The rotation period from TESS and chromospheric activity are similar to Pleiades members.

\medskip
\noindent{\bf  HIP 63734 = HD 113414}
F7/F8 star, investigated by Nardiello et al., submitted as a candidate comoving object to TOI-1807 and TOI-2048, 300 Myr old stars both hosting transiting planets.
The star results likely younger, with an estimated age of 150 Myr from Li EW and X-ray emission.
The identification of a faint BD candidate, compatible with the $\Delta \mu$ signature, is discussed in Sect. \ref{sec:tyc6539-HIP6734}.

\medskip
\noindent{\bf HIP 63862 = HD 113553 = TYC 8258-871-1}
BANYAN $\Sigma$ yields a 53\% membership probability for Carina-Near MG. The star was not considered a candidate members in past studies,
With a field star classification and a variety of age indicators, \citet{Vigan2017} estimated an age of 180 Myr with 120-250 limits, fully consistent with the age of Carina-Near MG.
We adopt Carina-Near MG age, with expanded age limits from indirect age indicators due to the 
moderately low membership probability.
Mean RVs from \citet{nordstrom2004,sacy,GaiaDR2} agree very well, while \citet{nordstrom2004} mentioned a marginally significant variability (rms 0.5 km/s, 11\% probability of constant RV).
The source at 4.4", previously detected by \citet{Chauvin2015}, is a background object.


\medskip
\noindent{\bf HIP 71899 = HD 129501 = TYC 1483-1030-1}
F8 star. A bright source is detected within IRDIS field of view and is also seen (without full astrometric solution) in \citet{GaiaDR2,GaiaEDR3}.
The combination of SPHERE and {\it Gaia}data confirms its physical association, with a mass of 0.39 Msun.
Indications of the young age of the system comes from the prominent X-ray emission (intermediate between Hyades and Pleiades of similar color) and 
rotation period from KELT \citep{oelkers2018}, comparable to Pleiades stars although with the limited age sensitivity expected for F type stars. Isochrone fitting, when adopting solar metallicity is compatible with the young age from activity and with a broader range up to 3 Gyr. The sub-solar metallicity by Stromgren photometry \citep{casagrande2011} (which however could be biased by stellar activity) would instead indicate an evolved star.
Finally the kinematics is slightly outside the kinematic space of young stars.
The limited amount of data (no light curve from TESS, no high resolution spectra available in public archives) and the impact of binarity prevents a better age determination.
Therefore, we adopt the X-ray age, keeping as upper limit the results of isochrone fitting.

\medskip
\noindent{\bf  HD 143600 = HIP 78549}
The star is a well known recognized member of Upper Sco association
\citep[e.g.,][]{dezeeuw1999} and membership is confirmed by our own check with {\it Gaia}data.
The star was discovered to be a close visual binary (sep 0.325", $\Delta$ Ks=3.99) by \citet{Lafreniere2014}.
There is no significant IR excess, while the star has a significant reddening (E(B-V) = 0.14) from comparison of observed and expected colors for the B9.5 spectral type. 

\medskip
\noindent{\bf  HIP 108912}
All the indicators are consistent with an age intermediate between Hyades and Pleiades, and similar to Group X. RV by \citet{nordstrom2004,GaiaDR2} are in agreement within errors.

\medskip
\noindent{\bf  HIP 112491}
Field object; age indicators consistently suggest an age close to that of the Hyades.

\medskip
\noindent{\bf  HIP 112581}
Field object with age similar to the Hyades.
A close candidate (estimated mass 0.14 Msun from photometry) is identified at 0.7", very likely a true companion considering the small projected separation and the presence of the astrometric signature, roughly compatible with the properties of the observed candidate.

\medskip
\noindent{\bf TYC 9339-2158-1}
Wide visual binary, with the primary of the system, the G1 star HIP 116063 = HD 221231 being at 36".
The analysis of the age indicators yields to an age intermediate between Hyades and Pleiades \citep{Vigan2017}.

\medskip
\noindent{\bf  HIP 116768}
Am star. Isochrone fitting \citep[using the Teff calibration by][]{netopil2008} indicate that the star is somewhat evolved outside main sequence. However, the star is likely a binary (RUWE=6.3 in {\it Gaia}EDR3) so parallax (which has a large error), magnitude, and age can be biased by binarity.
The star has a low membership probability in Argus (29.8\%), in contrast with the isochrone age.

\section*{Acknowledgements}
This research has made use of the SIMBAD database and of the VizieR catalogue access tool operated at CDS, Strasbourg, France, and of the Washington Double Star Catalogue maintained at the U.S. Naval Observatory.
This work has made use of data from the European Space Agency (ESA) mission {\it Gaia} (\url{http://www.cosmos.esa.int/gaia}), processed by the {\it Gaia} Data Processing and Analysis Consortium (DPAC, \url{http://www.cosmos.esa.int/web/gaia/dpac/consortium}). Funding for the DPAC has been provided by national institutions, in particular the institutions participating in the {\it Gaia} Multilateral Agreement.
This work has made use of the SPHERE Data Centre, jointly operated by OSUG/IPAG (Grenoble), PYTHEAS/LAM/CeSAM (Marseille), OCA/Lagrange (Nice), Observatoire de Paris/LESIA (Paris), and Observatoire de Lyon/CRAL, and supported by a grant from Labex OSUG@2020 (Investissements d’avenir – ANR10 LABX56).
CF acknowledges support from the Center for Space and Habitability (CSH). This work has been carried out within the framework of the NCCR PlanetS supported by the Swiss National Science Foundation.
This work has been supported by the PRIN-INAF 2019 "Planetary systems at young ages (PLATEA)" and ASI-INAF agreement n.2018-16-HH.0.
KM acknowledges funding by the Science and Technology Foundation of Portugal (FCT), grants No. IF/00194/2015, PTDC/FIS-AST/28731/2017, and UIDB/00099/2020.
For the purpose of open access, the authors have applied a Creative Commons Attribution (CC BY) licence to any Author Accepted Manuscript version arising from this submission.


\label{lastpage}
\end{document}